\documentclass[aip,graphicx]{revtex4-1}
\usepackage{bm}
\usepackage{graphicx}
\usepackage{amsmath}
\usepackage{amssymb}
\usepackage{color}
\usepackage{siunitx}
\usepackage{xr}
\usepackage{filecontents}
\newcommand{\rmd}{\mathrm{d}}
\newcommand{\kB}{k_\mathrm{B}}
\newcommand{\cH}{c_\mathrm{H}}
\newcommand{\cL}{c_\mathrm{L}}
\newcommand{\cinf}{c_\infty}

\newcommand{\pH}{p_\mathrm{H}}
\newcommand{\pL}{p_\mathrm{L}}
\newcommand{\pinf}{p_\infty}
\newcommand{\Dc}{\Delta c}
\newcommand{\Dp}{\Delta p}

\newcommand{\cs}{c_{\mathrm{s}}}

\newcommand{\Jstep}{J_\mathrm{step}}
\newcommand{\KDO}{\kappa_\mathrm{DO}}
\newcommand{\kDO}{K_\mathrm{DO}}
\newcommand{\Ps}{P_\mathrm{s}}

\newcommand{\lambdaD}{\lambda_\mathrm{D}}
\newcommand{\aeff}{a_\mathrm{eff}}
\newcommand{\csalt}{c_\mathrm{salt}}
\newcommand{\D}[2]{\frac{\partial #2}{\partial #1}}

\graphicspath{{./images/}}

\begin{document}
	
	% Use the \preprint command to place your local institutional report number 
	% on the title page in preprint mode.
	% Multiple \preprint commands are allowed.
	%\preprint{}
	
	\title{Entrance effects in concentration-gradient-driven flow through an ultrathin porous membrane} %Title of paper
	
	% repeat the \author .. \affiliation  etc. as needed
	% \email, \thanks, \homepage, \altaffiliation all apply to the current author.
	% Explanatory text should go in the []'s, 
	% actual e-mail address or url should go in the {}'s for \email and \homepage.
	% Please use the appropriate macro for the type of information
	
	% \affiliation command applies to all authors since the last \affiliation command. 
	% The \affiliation command should follow the other information.
	
	%\author{}
	%\email[]{Your e-mail address}
	%\homepage[]{Your web page}
	%\thanks{}
	%\altaffiliation{}
	%\affiliation{}
	
	\author{Daniel J. Rankin}
	\affiliation{Department of Chemistry, School of Physical Sciences, The University of Adelaide, Australia}
	
	\author{Lyd\'eric Bocquet}
	\affiliation{Laboratoire de Physique Statistique, CNRS UMR  8550, Ecole Normale Sup\'erieure, PSL Research University, Paris, France}
	
	\author{David M. Huang}
	\email{david.huang@adelaide.edu.au}
	%\phone{xxx}
	%\fax{xxx}
	\affiliation{Department of Chemistry, School of Physical Sciences, The University of Adelaide, Australia}
	
	% Collaboration name, if desired (requires use of superscriptaddress option in \documentclass). 
	% \noaffiliation is required (may also be used with the \author command).
	%\collaboration{}
	%\noaffiliation
	
	%\date{\today}

\begin{abstract}
Transport of liquid mixtures through porous membranes is central to processes such as desalination, chemical separations and energy harvesting, with ultrathin membranes made from novel 2D nanomaterials showing exceptional promise. Here we derive, for the first time, general equations for the solution and solute fluxes through a circular pore in an ultrathin planar membrane induced by a solute concentration gradient. We show that the equations accurately capture the fluid fluxes measured in finite-element numerical simulations for weak solute--membrane interactions. We also derive scaling laws for these fluxes as a function of the pore size and the strength and range of solute--membrane interactions. These scaling relationships differ markedly from those for concentration-gradient-driven flow through a long cylindrical pore or for flow induced by a pressure gradient or electric field through a pore in an ultrathin membrane. These results have broad implications for transport of liquid mixtures through membranes with a thickness on the order of the characteristic pore size.
\end{abstract}

%\pacs{}% insert suggested PACS numbers in braces on next line

\maketitle %\maketitle must follow title, authors, abstract and \pacs

% Body of paper goes here. Use proper sectioning commands. 
% References should be done using the \cite, \ref, and \label commands
%\section{}
%%\label{}
%\subsection{}
%\subsubsection{}

\section{Introduction}

Fluid transport through pores and porous membranes plays a key role in many processes of fundamental and practical interest, including cellular homeostasis in biological systems, \cite{Sui2001} chemical separations, \cite{Werber2016} desalination, \cite{Elimelech2011} and energy conversion. \cite{Logan2012,Siria2017} Thus, a general theoretical understanding of the parameters that control these transport phenomena has broad implications for a variety of domains. Many theoretical models of fluid transport in porous membranes have considered flows only within the pores \cite{Fair1971,Bonthuis2011,Balme2015,Peters2016} and have neglected the effect of transport between the membrane pores and the fluid outside the membrane. These so-called entrance or access effects can dominate fluid transport processes when the membrane thickness approaches the characteristic pore size \cite{Sherwood2014,Melnikov2017} or when the fluid--solid friction becomes small. \cite{Sisan2011,Rankin2016} The most extreme examples of this situation are membranes of atomic thickness made from 2D materials such as graphene and its derivatives \cite{Surwade2015,Cohen-Tanugi2016,Morelos-Gomez2017,Li2016d,Walker2017,Wang2017i} or molybdenum sulfide. \cite{Feng2016,Li2016a} Such 2D membranes are of great interest, as they have been shown to exhibit exceptional properties compared with conventional membranes for applications such as desalination\cite{Surwade2015} and electrical energy harvesting from salinity gradients. \cite{Feng2016}

Fluid fluxes across a membrane can be induced by a variety of driving forces, including gradients of pressure, electrical potential, or solute concentration. Equations have previously been derived to quantify entrance effects on fluid flow driven by a pressure gradient \cite{Sampson1891,Weissberg1962} and on fluid flow \cite{Mao2014} and ionic electrical currents \cite{Hall1975,Lee2012g} induced by an electric field acting on a electrolyte solution. However, to date, no theory has been developed to describe the entrance effects on fluid fluxes driven by concentration gradients and how they vary with relevant parameters. 

Fluid fluxes driven by concentration gradients are of particular relevance in applications such as chemical separations, \cite{Werber2016,Morelos-Gomez2017} desalination, \cite{Elimelech2011,Surwade2015,Cohen-Tanugi2016,Li2016d} and salinity-gradient-driven energy harvesting. \cite{Logan2012,Siria2013,Feng2016,Rankin2016,Siria2017} The work presented here focuses specificially on entrance effects on the concentration-gradient-driven process of diffusioosmosis, \cite{Anderson1989} in which flow of a solute-containing solution is driven by an osmotic-pressure gradient that develops in the inhomogeneous interfacial fluid layer induced by interactions of the fluid with the solid surfaces. Diffusioosmosis has been shown to play a key role in astonishing energy densities measured for salinity gradient energy harvesting in a nanotube membrane. \cite{Siria2013} Thus, entrance effects on this phenomenon are of considerable interest.       

Here we derive, for the first time, general equations to quantify the diffusioosmotic solution flux and solute flux of a dilute solution through a circular aperture in a 2D membrane as a function of the aperture size and the strength and range of the interactions between the solute and membrane surface. We verify the accuracy of the equations by comparison with finite-element numerical simulations. We go on to compare the scaling behavior predicted for concentration-gradient-driven flow through a circular aperture with those for other membrane geometries and driving forces and discuss the implications of these results for real systems. 

\section{Theory}

\subsection{Diffusioosmotic flow}
\begin{figure}%[tb!]
	\centering
	\includegraphics{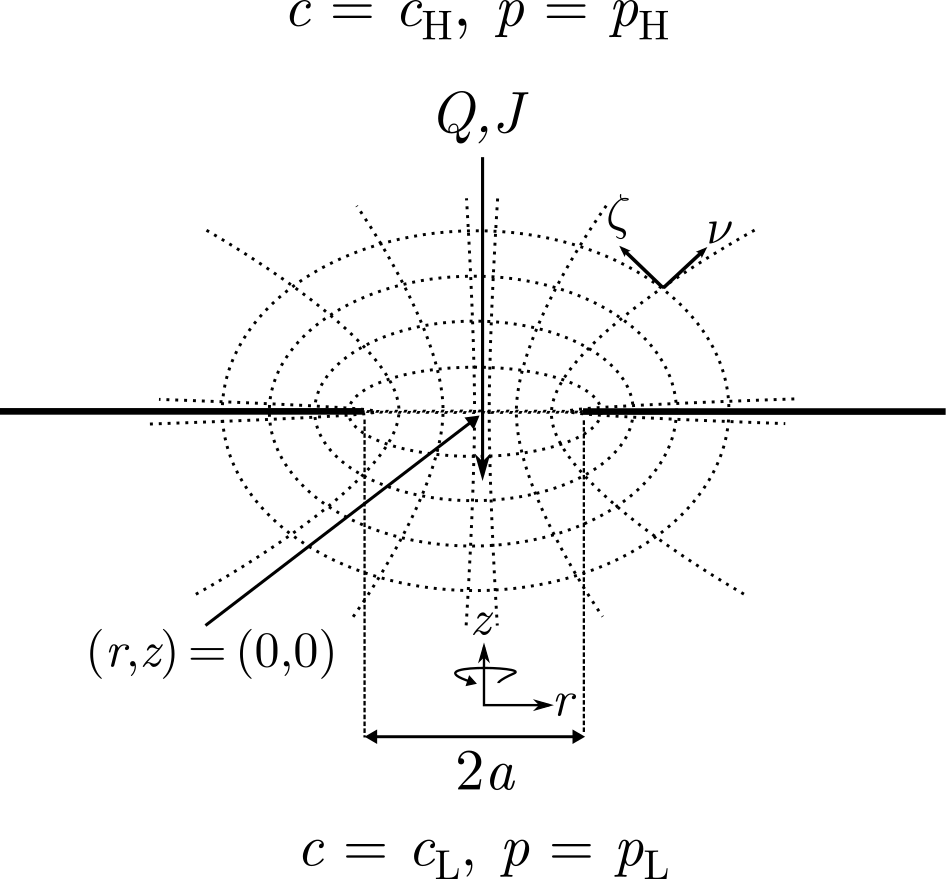}
	\caption{Schematic of flow of a solution through a circular aperture of radius $a$ in an infinitesimally thin planar wall. The origin in cylindrical $(r,z)$ coordinates is at the center of the aperture, as indicated, and the axis of symmetry is the depicted $z$ axis. The solute concentration $c$ and solution pressure $p$ far from the membrane are $\cH$ and $\pH$, respectively, in the upper half-plane and $\cL$ and $\pL$, respectively, in the lower half-plane. $Q$ and $J$ are the solution and solute fluxes, respectively. %For concentration-gradient-driven flow, $\pH = \pL$, whereas for pressure-driven flow, $\cH = \cL$. 
		Contours of constant $\zeta$ and $\nu$ in oblate--spheroidal $(\zeta,\nu)$ coordinates are also shown as dashed lines, with unit vectors shown at one point in space. 
		\label{fig:flow_geometry}
	}
\end{figure}

Consider the flow of a solution containing a single solute type through a circular aperture of radius $a$ in an infinitesimally thin planar wall, as illustrated in Fig.~\ref{fig:flow_geometry}. Assuming that the fluid flows can be described by continuum hydrodynamic equations for low-Reynolds-number steady-state flow of a dilute solution of an incompressible Newtonian fluid, the governing equations are \cite{Anderson1982}
\begin{align}
	-\nabla p - c\nabla U + \eta \nabla^2 \bm{u} &= 0, \label{eq:stokes}\\
	\nabla \cdot \bm{j} = \nabla \cdot \left(-D\nabla c - \mu c \nabla U + \bm{u}c\right) &= 0, \label{eq:delj}\\
	\nabla \cdot \bm{u} &= 0, \label{eq:delu}
\end{align}
where $\bm{u}$, $\bm{j}$, $p$, and $c$ are the solution velocity, solute current density, pressure, and solute concentration, respectively, $\eta$ is the solution shear viscosity, and $U$ is the solute--membrane interaction potential. $D$ and $\mu$ are the solute diffusivity and mobility respectively, which we assume are related by the Einstein relation, $\mu = \frac{D}{\kB T}$, \cite{Probstein1994} where $\kB$ is the Boltzmann constant and $T$ is the temperature.
$U$ is the interaction potential per solute molecule and so $-c\nabla U$ is a body force per unit volume acting on the fluid due to the solute--membrane interactions. $U$ is assumed to depend on the position in the fluid relative to the membrane surface. For a neutral solute, $U$ typically depends on the distance from the surface \cite{Anderson1989,Anderson1982} and has a range on the order of the solute molecular diameter.
Further assuming that advection of the solute is negligible compared with diffusion (i.e. low P\'eclet number flow), Eq.~\eqref{eq:delj} for the solute flux simplifies to 
\begin{equation}
	\nabla \cdot \bm{j} = \nabla \cdot \left(-D\nabla c - D\frac{c}{\kB T}\nabla U\right) = 0. 
	\label{eq:delj-lowPe}
\end{equation} 
The solution velocity and solute flux are assumed to satisfy the usual no-slip and zero flux boundary conditions at the membrane surface, i.e. $\bm{u} = 0$ and $\hat{\bm{n}}\cdot \bm{j} = 0$, where $\hat{\bm{n}}$ is the unit vector normal to the membrane surface.

We note that a similar approach based on the widely used Poisson--Nernst--Planck--Stokes equations for electrolytes, \cite{Probstein1994} in which the electric potential energy plays an analogous role for the electrolyte that the interaction potential $U$ does for a neutral solute, could be used to extend this study to concentration-gradient-driven electrolyte transport. However, such an extension is non-trivial, as the electric potential must be determined by solving an additional coupled differential equation (the Poisson equation) that depends on the solute (electrolyte) concentration, rather than being specified. Thus, we leave this extension to electrolytes to future work.

Consider the fluid flow induced by a concentration difference, $\Dc = \cH - \cL$, between the two sides of the membrane, with the pressure far from the membrane the same on both sides of the membrane, i.e. $\pH = \pL =\pinf$, as shown in Fig.~\ref{fig:flow_geometry}. 

Our derivation uses a combination of cylindrical $(r, z, \phi)$ and oblate spheroidal $(\zeta,\nu, \phi)$ coordinates, where  $z = a\nu \zeta$, $r = a \sqrt{(1+\nu^2)(1-\zeta^2)}$, and $\theta$ is the angle about the $z$ axis ($0\leq \nu < \infty$, $-1\leq \zeta \leq 1$, and $0\leq\phi< 2\pi$).  \cite{Morse1953-oblate-spheroidal,Mao2014} 
Writing the solute concentration in the presence of a concentration gradient as 
\begin{equation}
	c(\zeta,\nu) \equiv \cs(\zeta,\nu) e^{-U(\zeta,\nu)/\kB T},
	\label{eq:c}
\end{equation}
where $\cs(\zeta,\nu)$ is to be determined, and inserting this expression into Eq.~\eqref{eq:delj-lowPe} and the boundary condition for the solute current density $\bm{j}$ gives 
\begin{equation}
	\nabla^2 \cs -\nabla \left(\frac{U}{\kB T}\right) \cdot \nabla \cs = 0,
	\label{eq:nabla-cs}
\end{equation}
with the boundary condition $\hat{\bm{n}} \cdot \nabla \cs = 0$ at the membrane surface. If the solute--membrane interaction potential is small relative to the thermal energy ($U \ll \kB T$), Eq.~\eqref{eq:nabla-cs} reduces to
\begin{equation}
	\nabla^2 \cs = 0. 
	\label{eq:cs-laplace}
\end{equation} 
Solving this equation, subject to the boundary conditions on $\cs$ at the membrane surface and far from the membrane ($\cs \rightarrow \cH = \cinf + \frac{\Dc}{2}$ and $\cL = \cinf-\frac{\Dc}{2}$ in the upper and lower half-planes, respectively, where $U \rightarrow 0$), gives \cite{Morse1953-oblate-spheroidal}
\begin{equation}
	\cs = \cinf + \frac{\Delta c}{\pi}\tan^{-1}{\nu}.  
	\label{eq:cs}
\end{equation}
We have verified using finite-element numerical simulations (see Fig.~\ref{Sfig:solute_conc} of the supplementary material) that Eq.~\eqref{eq:cs} with Eq.~\eqref{eq:c} accurately describe the solute concentration even when $U$ is several times $\kB T$. A possible reason why Eq.~\eqref{eq:cs} appears to be accurate outside the regime for which it was derived is that the second term in Eq.~\eqref{eq:nabla-cs} that was neglected to arrive at Eq.~\eqref{eq:cs-laplace} and~\eqref{eq:cs} can be small even if the magnitude of the potential $U$ is large: for example, for $\cs$ given by Eq.~\eqref{eq:cs} this term is zero for a potential  that is a function only of $\zeta$ or in the pore mouth (at $\nu = z=0$) for a potential that is a function only of distance from the membrane, due to the orthogonality of $\nabla U$ and $\nabla \cs$ in these cases.

The fluid flow induced by the concentration gradient can be obtained from the reciprocal theorem for steady incompressible creeping flow, \cite{Happel1983} which allows the fluid flow due to a body force $\bm{F}$ acting on the fluid to be related to the pressure-driven flow in the same pore geometry, \cite{Mao2014} for which an analytical solution exists for the fluid velocity through a circular aperture. \cite{Sampson1891,Happel1983}  As shown by Mao et al. \cite{Mao2014} for the related problem of electroosmosis through a circular aperture,
\begin{equation}
	Q = -\frac{1}{\Dp}\iiint_V \rmd V\, \bar{\bm{u}}\cdot \bm{F},
	\label{eq:reciprocal}
\end{equation}
where $\bar{\bm{u}}$ is the fluid velocity induced by a pressure difference $\Dp = \pH -\pL $ for $\Dc = 0$ in the system geometry in Fig.~\ref{fig:flow_geometry} and the integral is over the volume $V$ occupied by the fluid. For concentration-gradient-driven flow described by Eq.~\eqref{eq:stokes}, $\bm{F} = -c\nabla U$. The pressure-driven flow velocity can be obtained from the stream function $\psi = -\frac{a^3}{6\pi\eta}\left(1-\zeta^3\right)\Dp$ for the flow \cite{Happel1983} using $\bar{\bm{u}} = \frac{1}{r}\hat{\bm{\phi}}\times\nabla\psi$, \cite{Happel1983} where $\hat{\bm{\phi}}$ is the unit vector in the $\phi$ direction, as 
\begin{equation}
	\bar{\bm{u}} = -\frac{a\zeta^2\Dp}{2\pi\eta\sqrt{(1+\nu^2)(\nu^2+\zeta^2)}}\;\hat{\bm{\nu}}. 
	\label{eq:u_pressure}
\end{equation}
Inserting this expression for $\bar{\bm{u}}$ and  $\bm{F} = -c\nabla U$ into Eq.~\eqref{eq:reciprocal} (with $c$ given by Eqs.~\eqref{eq:c} and~\eqref{eq:cs}) and making use of  
$\D{\nu}{\cs}=\frac{\Dc}{\pi}({1+\nu^2})^{-1}$
yields
\begin{equation}
	Q=-\KDO \Dc,
\end{equation}
with
\begin{equation}
	\KDO = -\frac{2\kB T a^3}{\pi\eta} \int_0^1 \rmd\zeta\,\zeta^2\int_0^\infty \rmd\nu\,\left(\frac{e^{-U/\kB T}-1}{1+\nu^2}\right). 
	\label{eq:Qzetanu}
\end{equation}
Equation~\eqref{eq:Qzetanu} is the main result of this work.

Furthermore, the solute flux density can be obtained, using Eqs.~\eqref{eq:delj-lowPe} and \eqref{eq:c}, as 
\begin{equation}
	\bm{j} = -D e^{-U/\kB T}\nabla \cs. 
	\label{eq:j}
\end{equation} 
The solute flux across the membrane is 
\begin{equation}
	J = \iint_S \rmd s\, \bm{j}\cdot \bm{\hat n}, 
	\label{eq:J}
\end{equation}
where the unit vector normal to the pore mouth is $\bm{\hat n}=\bm{\hat \nu}=\bm{\hat z}$ and the surface integral is over the pore aperture.
Evaluating the solute flux at the pore mouth (at $\nu = z = 0$) using Eqs~\eqref{eq:cs} and ~\eqref{eq:j} yields
\begin{align}
	J &= -2Da\Delta c \int_0^1 \rmd \zeta\, e^{-U/\kB T},  \label{eq:Jzeta}\\
	&= -2D\Delta c \int_0^a \frac{\rmd r\, re^{-U/\kB T}}{\sqrt{a^2-r^2}}. 
	\label{eq:Jr}
\end{align} 

\subsection{Limiting cases and scaling behavior}

The diffusioosmotic mobility predicted in Eq.~\eqref{eq:Qzetanu} depends crucially on the range  of the interaction potential $U$, which we will call $\lambda$. 
However,  the term $\exp(-U/\kB T)-1$ in Eq.~\eqref{eq:Qzetanu} is averaged spatially with a geometry-dependent weight, with a complicated dependence on the (oblate--spheroidal) coordinates $\zeta$ and $\nu$. As a consequence, the mobility, and its scaling with the pore radius $a$ and interaction range $\lambda$, may depend on  specific details of the geometry dependence of the interaction.
Therefore, we consider various limiting cases for the geometry of the potential and the consequences for the dependence of the scaling with the pore radius and interaction range.

\subsubsection{Case of potential that depends only on the distance to the membrane surface and/or pore edge}

In most circumstances, the potential $U$ is expected to be a function of the distance $d$ from the membrane surface. However,  the integral in Eq.~\eqref{eq:Qzetanu} for the mobility cannot be simplified due to the geometrical interplay between the various variables. 

Let us assume for simplicity that the solute excess/depletion at the membrane surface can be represented by a step function as a function of the distance $d$ from the surface, i.e.
\begin{equation}
e^{-U(d)/k_\mathrm{B}T} = \left\{ \begin{array}{ll}
\alpha, & d \leq \lambda \\
1, & d > \lambda \\
\end{array} \right. , \label{eq:Ustep}
\end{equation} 
where $\alpha$ characterizes the solute excess close to the membrane surface ($\alpha > 1$ for solute adsorption and $\alpha < 1$ for solute depletion).
While the mobility still cannot in general be evaluated analytically in this case, analytical solutions exists in certain limits.   
In particular, for a solute--wall interaction range much larger than the pore radius ($\lambda \gg a$), $e^{-U(d)/\kB T}$ can be approximated as a constant independent of the coordinates, and we find from Eq.~\eqref{eq:Qzetanu} in this limit that
\begin{equation}
\KDO \simeq -\frac{\kB T\Delta c}{3\eta}\left(\alpha - 1\right)a^3 , \quad \lambda \gg a. 
\label{eq:Qstep_big}
\end{equation}
On the other hand, if $\lambda \ll a$, the integral in Eq.~\eqref{eq:Qzetanu} can be approximated as
\begin{equation}
\KDO\simeq -\frac{\kB T}{\eta}\left(\frac{1}{4} + \frac{1}{\pi}\right) \left(\alpha - 1\right)a\, \lambda^{2}.
\label{eq:Qstep_small}
\end{equation}
Details of the derivation of this equation can be found in the supplementary material.

A similar calculation can be performed for the case in which the interaction originates only from the pore edge, and $U$ depends on the 
distance to the edge. As detailed in the supplementary material, the mobility in this case is
\begin{equation}
\KDO \simeq -\frac{\kB T}{2\eta}\left(\alpha - 1\right)a\, \lambda^{2}.
\end{equation}

\textit{Solute flux.} Similarly, using Eq.~\eqref{eq:Ustep} in Eq.~\eqref{eq:Jr} and noting that $d = a - r$ in the pore mouth at $z=0$, the solute flux across the membrane for the step-function potential can be shown to be
\begin{equation}
\Jstep = -2D\Dc \left[a+(\alpha-1)\sqrt{\lambda(2a-\lambda)}\right] 
\label{eq:Jstep}
\end{equation}
for $\lambda \leq a$ and 
\begin{equation}
\Jstep \simeq -2D\Dc \alpha a
\label{eq:Jstep-big}
\end{equation}
for $\lambda \gg a$.

\subsubsection{Case of potential that depends only on $\zeta$}

In the case in which the potential $U$ is a function of  the $\zeta$ coordinate only (see Fig.~\ref{fig:flow_geometry}),  the diffusioosmotic mobility in Eq.~\eqref{eq:Qzetanu} simplifies to 
\begin{equation}
	\KDO = -\frac{\kB T a^3\Dc}{\eta} \int_0^1 \rmd\zeta\,\zeta^2\left({e^{-U(\zeta)/\kB T}-1}\right). 
	\label{eq:Qzeta}
\end{equation}
Assuming that $U$ only depends on $\zeta$ is a stringent condition in terms of symmetry, but interestingly such a result is expected for a circular aperture in a planar membrane at a fixed electrostatic potential (however, in the absence of screening). 

To determine how $Q$ scales with $\lambda$, particularly in the limit $\lambda \ll a$, we can use the relationship between $\zeta$ and the distances, $d_1$ and $d_2$, from a point with this $\zeta$ value to the two foci of the hyperbolae or ellipses of constant $\zeta$ or $\nu$ that are shown in Fig.~\ref{fig:flow_geometry}. For an infinitesimally thin membrane, these foci are located at the pore edge (at $r=a$), and thus $1-\zeta^2 = \left[(d_1 - d_2)/(2a)\right]^2$. So a potential $U(\zeta)$ that depends only on $\zeta$ depends only on the relative distance, $d_1 - d_2$. Furthermore, assuming that the potential has a distance range $\lambda$ implies that
$U(\zeta) = U\left[\left(a-\vert d_1-d_2\vert/2\right)/\lambda\right]$. In addition, $U$ is only non-zero when $a-\vert d_1-d_2\vert /2 \sim \lambda \ll a$, so $a-\vert d_1-d_2\vert/2 = a(1-\sqrt{1-\zeta^2}) \simeq a \zeta^2/2$ when the integrand in Eq.~\eqref{eq:Qzeta} is non-zero. Thus, in this limit,  $U(\zeta) \sim U\left[a \zeta^2/(2\lambda)\right]$ and the integral in Eq.~\eqref{eq:Qzeta} becomes
\begin{equation}
	\int_0^1 \rmd\zeta\,\zeta^2\left({e^{-U\left[a \zeta^2/(2\lambda)\right]/\kB T}-1}\right) \simeq \left(2\lambda/a\right)^{3/2}\int_0^\infty \rmd x\,x^2 \left({e^{-U(x^2)/\kB T}-1}\right),
	\label{eq:Qzeta_scaling}
\end{equation}
where $x \equiv [a/(2\lambda)]^{1/2}\zeta$. So the mobility in this case scales as $\KDO \sim a^3(\lambda/a)^{3/2} \sim (a\lambda)^{3/2}$.

Equation~\eqref{eq:Qzeta} can also be rewritten in radial coordinates, by focusing on  the pore mouth ($\nu=z=0$),  as
\begin{equation}
	\KDO = -\frac{\kB T}{\eta}\int_0^a \rmd rr\sqrt{a^2-r^2}\left(e^{-U/\kB T}-1\right), \label{eq:Qr}
\end{equation} 
where $r$ is the distance to the center of the pore in the membrane plane ($z=0$). As an alternative approach to predicting the dependence of the mobility on $\lambda$, we again restrict ourselves for simplicity to a step-function interaction versus distance $d=a-r$ from the pore edge, i.e. 
\begin{equation}
	e^{-U(r)/k_\mathrm{B}T} = \left\{ \begin{array}{ll}
		\alpha, & a-r \leq \lambda \\
		1, & a-r> \lambda \\
	\end{array} \right. . \label{eq:Ustep0}
\end{equation} 
 Inserting Eq.~\eqref{eq:Ustep0} into Eq.~\eqref{eq:Qr} gives the diffusioosmotic mobility for the step-function potential as 
\begin{equation}
	\KDO = -\frac{\kB T}{3\eta}\left(\alpha - 1\right)\left[\lambda(2a-\lambda))\right]^{3/2}. 
	\label{eq:Qstep_small1}
\end{equation}
For a small interaction range $\lambda$ compared with the pore radius, the predicted scaling of the mobility is
\begin{equation}
	\KDO \simeq  -\frac{\kB T}{3\eta}\left(\alpha - 1\right)(a\lambda)^{3/2},  
\end{equation}
which is identical to the scaling predicted directly from Eq.~\eqref{eq:Qzeta}. 

\subsubsection{Case of potential that depends only on $\nu$}

In the case in which the potential $U$ is a function of  the $\nu$ coordinate only (see Fig.~\ref{fig:flow_geometry}),  the diffusioosmotic mobility in Eq.~\eqref{eq:Qzetanu} simplifies to 
\begin{equation}
	\KDO = -\frac{2\kB T a^3}{3\pi\eta} \int_0^\infty \rmd\nu\,\left(\frac{e^{-U(\nu)/\kB T}-1}{1+\nu^2}\right). 
	\label{eq:Qnu}
\end{equation}
Following similar reasoning to the previous section, in terms of the distances, $d_1$ and $d_2$, from a point with a given $\nu$ value to the foci of the ellipses or hyperbolae in Fig.~\ref{fig:flow_geometry}, $1+\nu^2 = \left[(d_1 + d_2)/(2a)\right]^2$. So a potential $U(\nu)$ that depends only on $\nu$ depends only on the average distance, $(d_1+d_2)/2$. Therefore, assuming that the potential has a distance range $\lambda$ entails $U(\nu)=U\left(\left[(d_1+d_2)/2 - a\right]/\lambda\right)$. For $\lambda \ll a$, $(d_1+d_2)/2 - a = a(\sqrt{1+\nu^2} - 1) \simeq a \nu^2 /2$. Thus, in this limit, $U(\nu) \sim U[a \nu^2 /(2\lambda)]$ and the integral in Eq.~\eqref{eq:Qnu} becomes
\begin{align}
	\int_0^\infty \rmd\nu\,\left(\frac{e^{-U[a \nu^2 /(2\lambda)]/\kB T}-1}{1+\nu^2}\right) &= (2\lambda/a)^{1/2}\int_0^\infty \rmd x\,\left(\frac{e^{-U(x^2)/\kB T}-1}{1+(2\lambda/a)x^2}\right), \label{eq:Qnu_scaling}\\
	&\simeq (2\lambda/a)^{1/2}\int_0^\infty \rmd x\,\left(e^{-U(x^2)/\kB T}-1\right),
	\label{eq:Qnu_scaling2}
\end{align}
where $x \equiv [a/(2\lambda)]^{1/2}\nu$. So the mobility in this case scales as $\KDO \sim a^3(\lambda/a)^{1/2} \sim a^{5/2}\lambda^{1/2}$.

\section{Numerical results}

To validate the theory, finite-element method (FEM) simulations of concentration-gradient-driven flow were carried out using Comsol Multiphysics (version 4.3a) \cite{Comsol4.3a} through pores with various radii and solute--membrane interactions. Here we consider that the solute interacts with the membrane via a potential that depends on the (shortest) distance $d$ to the membrane surface. Accordingly, the solute--membrane interaction potential was modelled using a hyperbolic tangent function, 
\begin{equation}
	U(d)=\frac{\epsilon}{2}\left[1-\tanh\left(\frac{d-\lambda}{\lambda}\right)\right], 
	\label{eq:Ufem}
\end{equation}
defined by parameters $\epsilon$ and $\lambda$ describing the strength and range of the potential. In all simulations, the average solute concentration $\cinf$ was \num{e-3}$\sigma^{-3}$ and the solute diffusivity $D$ was $\sigma^2/\tau$, and unless otherwise stated the aperture radius $a$ was $10~\sigma$ and $\lambda$ was $\sigma$, where $\sigma$ is the unit of length ($\sigma$ can be regarded as the diameter of a fluid molecule) and $\tau = \eta\sigma^3/(\kB T)$ is the unit of time. Details of the FEM simulations, which all correspond to low-P\'eclet number flow, are given in the supplementary material. 

We have quantified the concentration-gradient-driven solution and solute fluxes measured in the numerical simulations and predicted by our theory in terms of the diffusioosmotic mobility $\KDO$ defined by
\begin{equation}
	\KDO \equiv -\frac{Q}{\Dc}
	\label{eq:Gdef}
\end{equation}
and the solute permeance $\Ps$ defined by
\begin{equation}
	\Ps \equiv -\frac{J}{\Dc}.
	\label{eq:Gsdef}
\end{equation}
The equations that we have derived for the solution and solute fluxes (Eqs.~\eqref{eq:Qzetanu} and~\eqref{eq:Jzeta})  predict that the fluxes are linearly related to the concentration difference $\Dc$ and thus that the conductances and resistances defined in Eqs.~\eqref{eq:Gdef} and~\eqref{eq:Gsdef} are independent of $\Dc$. We have verified that this is indeed the case for the range of concentration differences studied in the numerical simulations ($\Dc = \num{e-6}$ to $\num{3e-4}~\sigma^{-3} = \num{e-3}$ to $\num{0.3}~\cinf$), as shown in Fig.~\ref{Sfig:flux_lin_resp} of the supplementary material. 

Figure~\ref{fig:do_mobility} shows the diffusioosmotic mobility $\KDO$ from the simulations and theory of a circular aperture as a function of aperture radius $a$ and solute--membrane interaction range $\lambda$ for two different values of the solute--membrane interaction strength $\epsilon$, $\kB T/10$ and $\kB T$, with all other parameters kept constant. The theory curves were calculated by numerically evaluating the integral in Eq.~\eqref{eq:Qzetanu} with the solute--membrane potential $U$ in Eq.~\eqref{eq:Ufem}. The sign of $\KDO$ has been defined so that a positive and negative values correspond to fluid flow in opposite and same direction, respectively, to the applied concentration gradient. Hence, for solute depletion at the membrane surface ($\epsilon > 0$) the flow is towards higher solute concentration ($\KDO < 0$), whereas for solute adsorption the flow is towards lower concentration ($\KDO > 0$). \cite{Huang2008c} 

\begin{figure}
	\centering
	\includegraphics{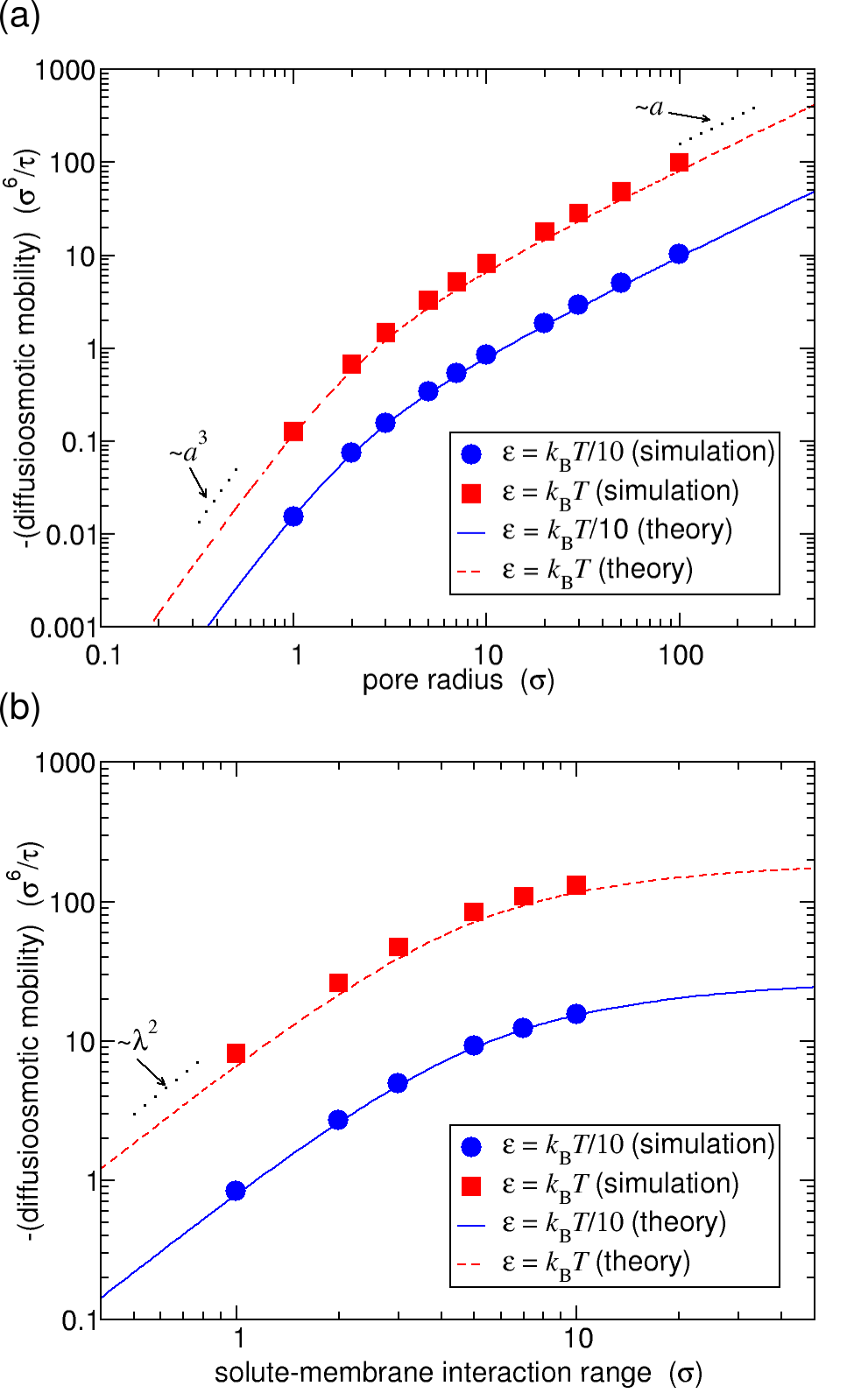}
	\caption{Diffusioosmotic mobility $\KDO$ versus (a) pore radius $a$ (with $\lambda = \sigma$) and (b) solute--membrane interaction range $\lambda$ (with $a = 10\sigma$) for solute--membrane interaction strength $\epsilon  = \kB T/10$ or $\kB T$ from FEM simulations (points) and theory (solid lines). The dashed lines show scaling with various powers of $a$ and $\lambda$.
		\label{fig:do_mobility}
	}
\end{figure}

Figure~\ref{fig:do_mobility} shows good quantitative agreement between the theory and simulations for the variation of $\KDO$ with all relevant parameters for $\epsilon \lesssim \kB T$. Since the theory assumes a weak potential in deriving Eq.~\eqref{eq:cs} for the solute concentration, %%that $U \ll \kB T$, 
we indeed find deviations between the prediction and the simulations as the magnitude of the solute--membrane potential increases.  Nevertheless, the agreement is reasonable well beyond the regime of validity of this approximation. Note that for the values of $\epsilon$ in Fig.~\ref{fig:do_mobility}, $\alpha - 1 \approx e^{-\epsilon/\kB T} - 1 \approx -\epsilon/\kB T$, and so from Eq.~\eqref{eq:Qstep_big} or~\eqref{eq:Qstep_small} $\KDO$ is approximately proportional to $\epsilon$, but this scaling is not expected in general and already starts to break down for $\epsilon = \kB T$.

Figure~\ref{fig:do_mobility} also compares the simulation results with the approximate scaling with the pore radius $a$ and solute--membrane interaction range $\lambda$ predicted by the theory. For $a \ll \lambda$, Eq.~\eqref{eq:Qstep_big} predicts that $\KDO$ is proportional to $a^3$ and independent of $\lambda$, which is evident in the scaling for small $a$ in  Fig.~\ref{fig:do_mobility}(a) and in the saturation at large $\lambda$ in Fig.~\ref{fig:do_mobility}(b). 
On the other hand, for $a \ll \lambda$, Eq.~\eqref{eq:Qstep_small} predicts scaling of $\KDO$ with $a\lambda^{2}$, which is seen to hold  in the large-$a$ regime in Fig.~\ref{fig:do_mobility}(a) and in the small-$\lambda$ regime in Fig.~\ref{fig:do_mobility}(b). 

\begin{figure}
	\centering
	\includegraphics{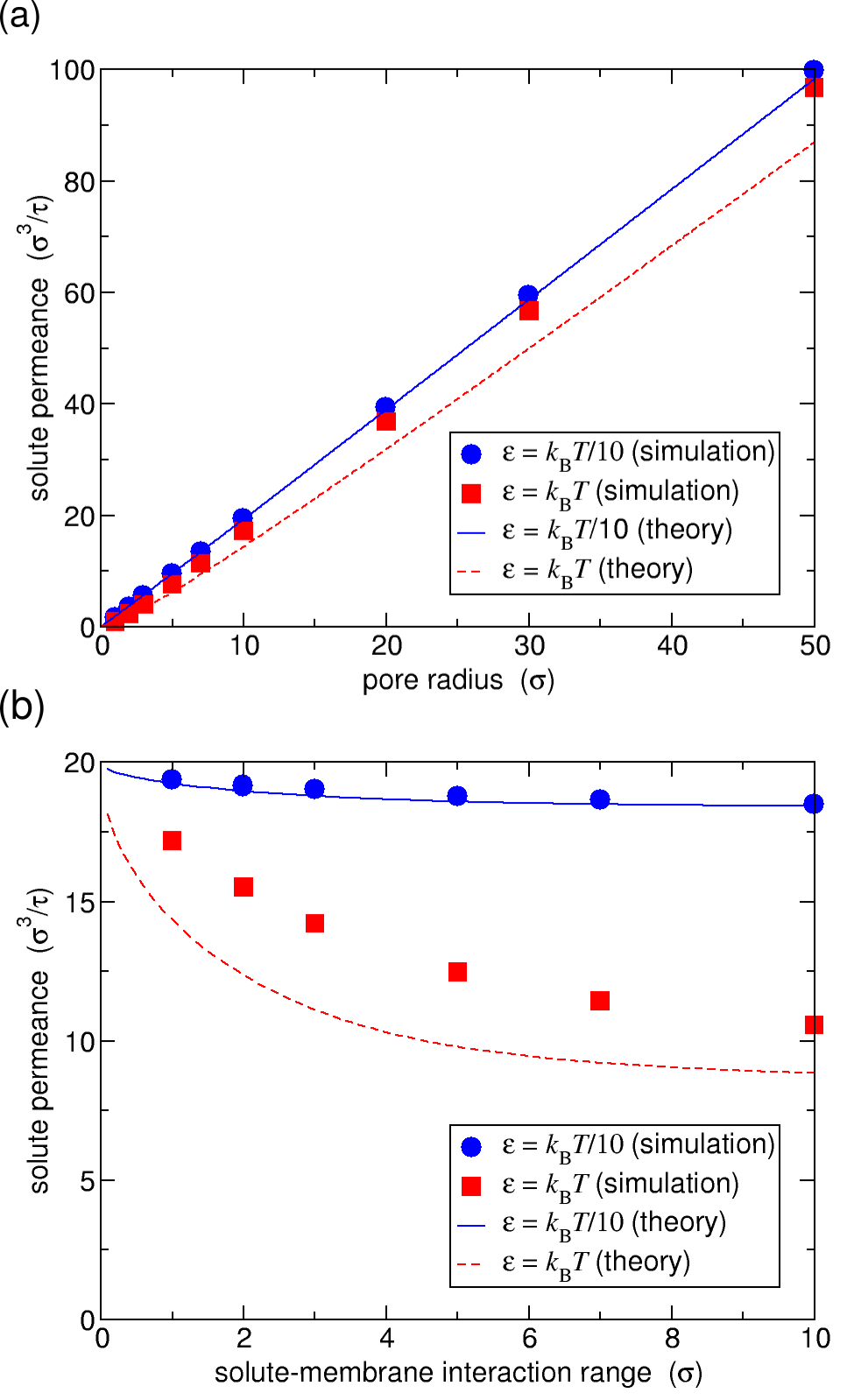}
	\caption{Solute permeance $\Ps$ versus (a) pore radius $a$ (with $\lambda = \sigma$) and (b) solute--membrane interaction range $\lambda$ (with $a = 10\sigma$) for solute--membrane interaction strength $\epsilon  = \kB T/10$ or $\kB T$ from FEM simulations (points) and theory (solid lines).
		\label{fig:solute_permeance}
	}
\end{figure}

Figure~\ref{fig:solute_permeance} shows the analogous comparison between the FEM simulations and theory for the solute permeance $\Ps$. The theory curves were calculated by numerically evaluating the integral in Eq.~\eqref{eq:Jr} with the solute--membrane potential $U$ in Eq.~\eqref{eq:Ufem}. As for the diffusioosmotic conductance, the theory accurately captures the simulated solute permeance for $\epsilon \lesssim \kB T$, with deviations between the theory and simulations becoming evident for magnitudes of the solute--membrane potential greater than $\kB T$. For the parameters used in Fig.~\ref{fig:solute_permeance}(a), the first term in Eq.~\eqref{eq:Jstep} dominates and so the permeance $\Ps$ shows the expected linear scaling with the pore radius $a$. For $\lambda \ll a$, Eq.~\eqref{eq:Jstep} predicts that $\Ps$ varies from its value at $\lambda = 0$ with a scaling as $\sqrt{\lambda}$, which is evident in Fig.~\ref{fig:solute_permeance}(b).

\section{Discussion}

An interesting outcome of the previous results is that the diffusioosmotic mobility $\KDO=-Q/\Delta c$ across a circular aperture in ultrathin membrane is strongly dependent on the details of the interaction of the solute with the membrane. We have shown in particular that the mobility scales with the pore radius $a$ and interaction range $\lambda$ as $a^{3-\gamma}\lambda^\gamma$, with an exponent $\gamma$ that depends on the underlying symmetries of the potential: when $\lambda \ll a$, for a potential that depends only on the $\zeta$ coordinate (in the oblate-spheroidal system, see Fig.~\ref{fig:flow_geometry}) an exponent $\gamma=3/2$ is predicted, for a potential that depends only on the $\nu$ coordinate an exponent $\gamma=1/2$ is predicted, whereas for a potential depending on the distance to the membrane a value $\gamma=2$ is found.

It is also interesting to compare the results for the circular aperture with those obtained in long cylindrical pores (e.g. as a model for nanotubes). As derived in detail in the supplementary material, the diffusioosmotic mobility of a long cylindrical pore of length $L$ is proportional to $(a\lambda)^{2}/L$ for $\lambda \ll a$. When compared to the case leading to an exponent $\gamma=2$, the scaling of $\KDO$ with pore size and interaction range $\lambda$ is therefore recovered by replacing the length of the nanopore $L$ with the pore size $a$, which is indeed expected for entrance effects. However, as shown with the case leading to an exponent $\gamma=3/2$ or $1/2$ for the diffusioosmotic mobility, this situation is not universal. The different scaling relationships derived for $\KDO$ for a circular aperture and a long cylindrical pore are summarized in Table~\ref{tab:scaling}.

 \begin{table}
 \caption{\label{tab:scaling}Scaling of the diffusioomotic mobility $\KDO$ with pore radius $a$ and solute--membrane interaction range $\lambda$ for a circular aperture in a 2D membrane, as well as with pore length $L$ for a long cylindrical pore, for different functional forms of the solute--membrane potential $U$ in the limits $\lambda \ll a$ and $\lambda \gg a$.}
 \begin{tabular}{cccc}
 	\hline\hline
 	System & Limit & Potential & Scaling\\
 	\hline
 	Circular aperture & $\lambda \ll a$ & $U(d)$ & $a\lambda^2$ \\
 	                  & $\lambda \ll a$ & $U(\zeta)$ & $a^{3/2}\lambda^{3/2}$ \\
 	 	              & $\lambda \ll a$ & $U(\nu)$ & $a^{5/2}\lambda^{1/2}$\\
 	 	              & $\lambda \gg a$ & any    & $a^3$ \\
 	\hline   
    Cylinder          & $\lambda \ll a$ & $U(d)$ & $a^2\lambda^2/L$\\	
                      & $\lambda \gg a$ & any      & $a^4/L$ \\
    \hline\hline 	 	 	                         
 \end{tabular}
 \end{table}

These result are relevant for transport through finite-length pores, for which the total resistance to flow can often be accurately given by the sum of the resistance due to the pore interior and that due to the pore ends, which can be approximated by that of a circular aperture. \cite{Weissberg1962,Lee2012g,Rankin2016} 

The predicted scaling behavior of the diffusioosmotic mobility and solute permeance of a circular aperture for concentration-gradient-driven flow differ markedly from the scaling behavior derived  previously for other types of flows in the same system geometry. For example, the hydraulic conductance (solution flux per unit pressure difference) in pressure-driven fluid flow through a circular aperture has been shown\cite{Sampson1891,Weissberg1962,Happel1983} to be proportional to $a^3$ in contrast to the proportionalty with $a^{3-\gamma}$ where $\gamma > 0$ for the diffusioosmotic mobility in concentration-gradient-driven flow in the limit $\lambda \ll a$.

On the other hand, for $\lambda \gg a$, the diffusioosmotic mobility shows the same $a^3$ scaling as the hydraulic conductance. The equivalent scaling of the hydraulic conductance and diffusioosmotic mobility in this limit is because concentration-driven-flow in which the solute--membrane interaction range is larger than the aperture radius is equivalent to osmotic transport through a semipermeable membrane, with the osmotic pressure gradient due to the concentration gradient playing an equivalent role to the pressure gradient in pressure-driven flow. \cite{Marbach2017}  Likewise, the scaling behavior predicted here for the diffusioosmotic mobility differs from the electroosmotic conductance for electric-field-driven fluid flow of an electrolyte, which has been shown to be proportional to $a^3/\lambdaD$ for $a \ll \lambdaD$, \cite{Mao2014} where $\lambdaD$ is the Debye length characterizing the electric double layer width, equivalent to $\lambda$ here; in the same limit, scaling of the diffusioosmotic mobility with $a^3$ is predicted here. 

The electrical conductance across a circular aperture in electric-field-driven transport of an electrolyte has been shown to be proportional to $a$ for an uncharged membrane, \cite{Hall1975} with the addition of surface charge to the membrane only changing the length scale in the scaling relationship to an effective radius $\aeff$ that is the sum of the aperture radius and the Dukhin length characterizing the ratio of the surface electrical conductivity to the bulk electrical conductivity, \cite{Lee2012g} without changing the scaling exponent. This scaling differs from that derived for the solute permeance for $\lambda \leq a$, for which the equivalent effective radius is $\aeff = \left[a+(\alpha-1)\sqrt{\lambda(2a-\lambda)}\right]$, in which the second term in the sum depends on both the solute--membrane interaction range $\lambda$ and pore radius $a$.

We can consider the implications of our theory for realistic systems, in particular for diffusioosmotic transport of electrolyte solutions. Extending the theory to electrolytes is desirable for applications, such as salinity-gradient-driven energy conversion\cite{Logan2012,Siria2013,Feng2016} and desalination, \cite{Elimelech2011,Surwade2015,Cohen-Tanugi2016,Li2016d} but it is technically difficult, so we leave this derivation for future work. However, as a rule of thumb, one may expect that this would amount to replacing the solute--membrane interaction range $\lambda$ by the salt-concentration-dependent Debye length $\lambdaD \propto \csalt^{-1/2}$. A counterintuitve outcome of the non-universal dependence of the diffusioosmotic mobility as a function of the interaction range $\lambda$ is a possible impact on the  dependence of the mobility on the salt concentration. 
In a long pore under a salinity difference $\Delta \csalt$, the solvent flux is predicted to behave as $Q = \kDO \Delta \log \csalt$, with the mobility $\kDO \propto \csalt^{0} \pi a^2/L$ for a long pore of length $L$, \cite{Siria2013}  i.e. no dependence on salt concentration. 
This is due to the dependence of the diffusioosmotic mobility $\KDO \sim \lambda_D^2\sim \csalt^{-1}$, so that $Q\sim \Delta \log \csalt$.
Now, for a circular aperture, we have shown that the dependence of the diffusioosmotic mobility $\KDO$ on the interaction range scales as 
$\KDO \sim \lambda^\gamma$ with a non-universal exponent $\gamma$. Thus, the case of $\gamma=2$ (occurring for potentials that depend on the distance to the membrane) will lead to $\kDO \propto \csalt^{0}$ as for the long-pore case; but cases with $\gamma\ne 2$ (as highlighted above in two cases), will lead to  $\kDO \propto \csalt^{1-\gamma/2}$, exhibiting therefore a curious dependence on $\csalt$. 

\section{Conclusions}

In summary, we have derived general equations and scaling relationships as a function of pore radius and solute--membrane interaction strength and range for the solution and solute fluxes induced by a solute concentration gradient through a circular aperture in an ultrathin planar membrane. 
We have shown, by comparing with finite-element numerical simulations, that the equations accurately quantify the fluid fluxes when the solute--membrane interaction strength is small compared with the thermal energy $\kB T$. 
In the limit of a solute--membrane interaction range much smaller than the pore radius, the theory predicts a non-universal  dependence of the fluid fluxes on the pore radius and interaction range. These results have significant implications for applications involving concencentration-gradient-driven flow in membranes in which the thickness is on the order of the pore size, such as those made from 2D nanomaterials, notably in the context of blue energy harvesting. 

\section*{Supplementary material}

The supplementary material contains derivations of scaling laws for diffusioosmosis through a circular aperture in an ultrathin planar membrane, the theory of concentration-gradient-driven flow in a long cylindrical pore, and further details and supplementary results of finite-element numerical simulations.

% If in two-column mode, this environment will change to single-column format so that long equations can be displayed. 
% Use only when necessary.
%\begin{widetext}
%$$\mbox{put long equation here}$$
%\end{widetext}

% Figures should be put into the text as floats. 
% Use the graphics or graphicx packages (distributed with LaTeX2e).
% See the LaTeX Graphics Companion by Michel Goosens, Sebastian Rahtz, and Frank Mittelbach for examples. 
%
% Here is an example of the general form of a figure:
% Fill in the caption in the braces of the \caption{} command. 
% Put the label that you will use with \ref{} command in the braces of the \label{} command.
%
% \begin{figure}
% \includegraphics{}%
% \caption{\label{}}%
% \end{figure}

% Tables may be be put in the text as floats.
% Here is an example of the general form of a table:
% Fill in the caption in the braces of the \caption{} command. Put the label
% that you will use with \ref{} command in the braces of the \label{} command.
% Insert the column specifiers (l, r, c, d, etc.) in the empty braces of the
% \begin{tabular}{} command.
%
% \begin{table}
% \caption{\label{} }
% \begin{tabular}{}
% \end{tabular}
% \end{table}

% If you have acknowledgments, this puts in the proper section head.
\begin{acknowledgments}
D.J.R. acknowledges the support of an Australian Government Research Training Program Scolarship and a University of Adelaide Faculty of Sciences Divisional Scholarship. L.B. acknowledges funding from the EU H2020 Framework Programme/ERC Advanced Grant agreement number 785911--Shadoks and ANR project Neptune.
\end{acknowledgments}

% Create the reference section using BibTeX:
\bibliography{ms}

% cross reference with supplementary information document
\makeatletter\@input{sup-xr.tex}\makeatother

\end{document}

	% Use the \preprint command to place your local institutional report number 
	% on the title page in preprint mode.
	% Multiple \preprint commands are allowed.
	%\preprint{}
	
	\title{Supplementary Material: Entrance effects in concentration-gradient-driven flow through an ultrathin porous membrane} %Title of paper
	
	% repeat the \author .. \affiliation  etc. as needed
	% \email, \thanks, \homepage, \altaffiliation all apply to the current author.
	% Explanatory text should go in the []'s, 
	% actual e-mail address or url should go in the {}'s for \email and \homepage.
	% Please use the appropriate macro for the type of information
	
	% \affiliation command applies to all authors since the last \affiliation command. 
	% The \affiliation command should follow the other information.
	
	%\author{}
	%\email[]{Your e-mail address}
	%\homepage[]{Your web page}
	%\thanks{}
	%\altaffiliation{}
	%\affiliation{}
	
	\author{Daniel J. Rankin}
	\affiliation{Department of Chemistry, School of Physical Sciences, The University of Adelaide, Australia}
	
	\author{Lyd\'eric Bocquet}
	\affiliation{Laboratoire de Physique Statistique, CNRS UMR  8550, Ecole Normale Sup\'erieure, PSL Research University, Paris, France}
	
	\author{David M. Huang}
	\email{david.huang@adelaide.edu.au}
	%\phone{xxx}
	%\fax{xxx}
	\affiliation{Department of Chemistry, School of Physical Sciences, The University of Adelaide, Australia}
	
	% Collaboration name, if desired (requires use of superscriptaddress option in \documentclass). 
	% \noaffiliation is required (may also be used with the \author command).
	%\collaboration{}
	%\noaffiliation
	
	%\date{\today}
	
	%\begin{abstract}
	% insert abstract here
	%\end{abstract}
	
	%\pacs{}% insert suggested PACS numbers in braces on next line

\maketitle %\maketitle must follow title, authors, abstract and \pacs

% Body of paper goes here. Use proper sectioning commands. 
% References should be done using the \cite, \ref, and \label commands
%\section{}
%\label{}
%\subsection{}
%\subsubsection{}

\newpage
\section{Derivation of scaling laws for diffusioosmosis through a circular aperture in a 2D planar membrane\label{Ssec:theory_orifice}}

Here we derive scaling laws for the diffusioosmotic flux through a circular aperture in an infinitesimally thin planar wall in the limit that the solute--membrane interaction range $\lambda$ is much smaller than the aperture radius $a$, i.e. $\lambda \ll a$. To derive analytical expressions, we assume a step-function form for the potential given by Eq.~\eqref{eq:Ustep} of the main paper. Inserting this equation into Eq.~\eqref{eq:Qzetanu} of the main paper for the diffusioosmotic flux gives
\begin{equation}
Q = -\frac{2\kB T a^3\Dc}{\pi\eta}\left(\alpha - 1\right) \int_0^1 \rmd\zeta\,\zeta^2\int_0^\infty \rmd\nu\,\frac{H\left[\lambda - d\left(\zeta,\nu\right)\right]}{1+\nu^2}, 
\label{Seq:Q_heaviside}
\end{equation}
where $\alpha$, as discussed in the main paper, is a parameter characterizing the solute excess close the membrane surface, $d \ge 0$ is the distance from the surface, and $H$ is the Heaviside step function, which sets the condition that the integrand is only non-zero when $0 \le d \le \lambda$. 

\subsection{Solute interaction with entire membrane surface\label{Ssec:theory_orifice_all}}

Using the relationships between the cylindrial $(r,z)$ and oblate--spheroidal $(\zeta,\nu)$ coordinates, $r = a\sqrt{(1+\nu^2)(1-\zeta^2)}$ and $z = a\nu\zeta$, gives
\begin{align}
d &= \left\{ \begin{array}{ll}
\sqrt{(a-r)^2+z^2}, & r \leq a \\
z, & r > a \\
\end{array} \right. , \label{Seq:d_nuzeta}\\
&= \left\{ \begin{array}{ll}
a\sqrt{2+\nu^2-\zeta^2-2\sqrt{(1+\nu^2)(1-\zeta^2)}}, & \sqrt{(1+\nu^2)(1-\zeta^2)} \leq 1 \\
a\nu\zeta, &  \sqrt{(1+\nu^2)(1-\zeta^2)} > 1 \\
\end{array} \right. .  \label{Seq:d_nuzeta2}
\end{align}
Since the integrand in Eq.~\eqref{Seq:Q_heaviside} is only non-zero when $d\le \lambda$, we only need to consider the case when this condition is satisfied to evaluate Eq.~\eqref{Seq:Q_heaviside}.    If $d\le \lambda \ll a$, then $\zeta \ll 1$ and $\nu \ll 1$ when $\sqrt{(1+\nu^2)(1-\zeta^2)} \leq 1$. In this case, using $\sqrt{1+\nu^2} = 1 + \nu^2/2 - \nu^4/8 + \ldots $  and $\sqrt{1-\zeta^2} = 1 - \zeta^2/2 - \zeta^4/8 - \ldots $, Eq.~\eqref{Seq:d_nuzeta2} simplifies to
\begin{equation}
d \simeq \left\{ \begin{array}{ll}
\frac{a}{2}\left(\nu^2 + \zeta^2 \right), & \nu \leq \zeta\\
a\nu\zeta, &  \nu > \zeta \\
\end{array} \right. .  \label{Seq:d_large_a}
\end{equation}
From this equation, for $d\ll a$, 
\begin{align}
\int_0^1 \rmd\zeta\,\zeta^2 &\int_0^\infty \rmd\nu\,
\frac{H\left[\lambda - d\left(\zeta,\nu\right)\right]}{1+\nu^2}  \nonumber \\ 
&\simeq \int_0^1 \rmd\zeta\,\zeta^2
\left\{
\int_0^\zeta \rmd\nu\,\frac{H\left[\lambda - a\left(\nu^2 + \zeta^2\right)/2\right]}{1+\nu^2} 
+ \int_\zeta^\infty \rmd\nu\,\frac{H\left(\lambda - a\nu\zeta\right)}{1+\nu^2} 
\right\}, \label{Seq:Qint_large_a}\\
&= \int_0^{\sqrt{\lambda/a}}  \rmd\zeta\,\zeta^2 \int_0^\zeta \frac{\rmd\nu}{1+\nu^2}
+ \int_{\sqrt{\lambda/a}}^{\sqrt{2\lambda/a}}  \rmd\zeta\,\zeta^2 \int_0^{\sqrt{2\lambda/a-\zeta^2}} \frac{\rmd\nu}{1+\nu^2}\nonumber\\
&\qquad + \int_0^{\sqrt{\lambda/a}}  \rmd\zeta\,\zeta^2 \int_\zeta^{\frac{\lambda}{a\zeta}} \frac{\rmd\nu}{1+\nu^2}, 
\label{Seq:Qint_large_a2}\\
&= \int_0^{\sqrt{\lambda/a}}  \rmd\zeta\,\zeta^2 \tan^{-1}\left(\frac{\lambda}{a\zeta}\right)
+ \int_{\sqrt{\lambda/a}}^{\sqrt{2\lambda/a}}  \rmd\zeta\,\zeta^2 \tan^{-1}\left(\sqrt{2\lambda/a-\zeta^2}\right), 
\label{Seq:Qint_large_a3}\\
&= -\frac{1}{12}\left[\frac{\left(a+3\lambda\right)\pi + 2\lambda}{a}\right] +\frac{\pi}{6}\left(\frac{a+2\lambda}{a}\right)^{3/2}
+\frac{1}{6}\left(\frac{\lambda}{a}\right)^2\nonumber\\
&\qquad - \frac{1}{3}\left(\frac{a+2\lambda}{a}\right)^{3/2}\tan^{-1}\left(\sqrt{\frac{a}{a+2\lambda}}\right)
-\frac{1}{6}\left(\frac{\lambda}{a}\right)^3\ln\left(1+\frac{a}{\lambda}\right).
\label{Seq:Qint_large_a4}
\end{align}
Inserting Eq.~\eqref{Seq:Qint_large_a4} into Eq.~\eqref{Seq:Q_heaviside} yields
\begin{align}
Q &\simeq -\frac{\kB T\Dc}{\eta}\left(\alpha - 1\right)
\left\{ 
-\frac{\left[\left(a+3\lambda\right)\pi + 2\lambda\right]a^2}{6\pi} +\frac{\left(a+2\lambda\right)^{3/2}a^{3/2}}{3}
+\frac{1}{3\pi}a\lambda^2 \right.  \nonumber\\
&\hspace*{22ex} \left.
-\frac{2}{3\pi}\left(a+2\lambda\right)^{3/2}a^{3/2}\tan^{-1}\left(\sqrt{\frac{a}{a+2\lambda}}\right)
-\frac{1}{3\pi}\lambda^3\ln\left(1+\frac{a}{\lambda}\right)
\right\}, 
\label{Seq:Q_large_a}\\
&=  -\frac{\kB T\Dc}{\eta}\left(\alpha - 1\right)a^3
\left\{
\left(\frac{1}{4} + \frac{1}{\pi} \right)\left(\frac{\lambda}{a}\right)^2 
+ \left(\frac{-4-3\pi-12\ln a + 12\ln \lambda}{36\pi}\right)\left(\frac{\lambda}{a}\right)^3  %\left(\frac{-16+3\pi}{48\pi}\right)\frac{\lambda^4}{a} 
+ O\left[\left(\frac{\lambda}{a}\right)^4 \right]
\right\}.  \label{Seq:Q_large_a_series}
\end{align}
Thus, for $\lambda \ll a$,
\begin{equation}
Q \simeq -\frac{\kB T\Dc}{\eta}\left(\frac{1}{4} + \frac{1}{\pi} \right)\left(\alpha - 1\right)a\lambda^2.  
\end{equation}

\subsection{Solute interaction with pore mouth only\label{Ssec:theory_orifice_mouth}}

Assuming that the solute interacts only with the part of the membrane surface at the pore mouth, the solute--membrane potential $U(d)$ will only depend on the distance from the surface at radial coordinate $r = a$. In this case, the distance is $d = \sqrt{(a-r)^2+z^2} , \forall r,z$. Thus, following analogous steps to those in the previous section, if $d\le \lambda \ll a$, $d \simeq \frac{a}{2}\left(\nu^2 + \zeta^2 \right)$ and
\begin{align}
\int_0^1 \rmd\zeta\,\zeta^2 \int_0^\infty \rmd\nu\,
\frac{H\left[\lambda - d\left(\zeta,\nu\right)\right]}{1+\nu^2}
&\simeq \int_0^1 \rmd\zeta\,\zeta^2
\int_0^\infty \rmd\nu\,\frac{H\left[\lambda - a\left(\nu^2 + \zeta^2\right)/2\right]}{1+\nu^2},  \label{Seq:Qint_mouth_large_a}\\
&=  \int_0^{\sqrt{2\lambda/a}}  \rmd\zeta\,\zeta^2 \int_0^{\sqrt{2\lambda/a-\zeta^2}} \frac{\rmd\nu}{1+\nu^2},  \label{Seq:Qint_mouth_large_a2}\\
&=  \int_0^{\sqrt{2\lambda/a}}  \rmd\zeta\,\zeta^2 \tan^{-1}\left(\sqrt{2\lambda/a-\zeta^2}\right) 
\label{Seq:Qint_mouth_large_a3}\\
&= \frac{\pi}{6}\left[\left(\frac{a+2\lambda}{a}\right)^{3/2}
-\left(\frac{a+3\lambda}{a}\right)
\right].
\label{Seq:Qint_mouth_large_a4}
\end{align}
Inserting Eq.~\eqref{Seq:Qint_mouth_large_a4} into Eq.~\eqref{Seq:Q_heaviside} yields
\begin{align}
Q &\simeq -\frac{\kB T\Dc}{3\eta}\left(\alpha - 1\right)
\left[\left(a+2\lambda\right)^{3/2}a^{3/2}-\left(a+3\lambda\right)a^2\right],
\label{Seq:Q_mouth_large_a}\\
&=  -\frac{\kB T\Dc}{\eta}\left(\alpha - 1\right)a^3
\left\{\frac{1}{2}\left(\frac{\lambda}{a}\right)^2 
- \frac{1}{6}\left(\frac{\lambda}{a}\right)^3 + O\left[\left(\frac{\lambda}{a}\right)^4 \right] \right\}.  \label{Seq:Q_mouth_large_a_series}
\end{align}
Thus, for $\lambda \ll a$,
\begin{equation}
Q \simeq -\frac{\kB T\Dc}{2\eta}\left(\alpha - 1\right)a\lambda^2.  
\end{equation}
Thus, the scaling of $Q$ with the pore radius $a$ and solute--membrane interaction range $\lambda$ and strength (measured by $\alpha$) is the same as in the case of solute interactions with the entire membrane.

\section{Theory of concentration-gradient-driven flow in a long cylindrical pore\label{Ssec:theory_long_pore}}

\begin{figure}[tb!]
	\centering
	\includegraphics[width=9cm]{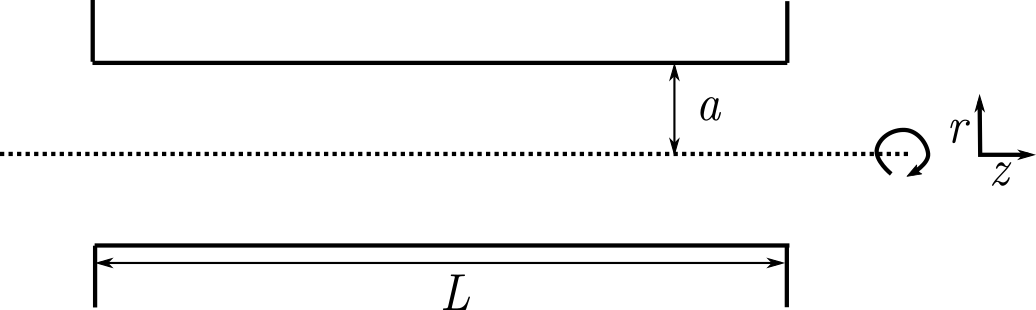}
	\caption{Cylindrical pore geometry for deriving fluxes through a long pore. The pore has a radius of $a$ and a length of $L$. The dashed line indicates the axis of symmetry.
		\label{Sfig:cylinder}
	}
\end{figure}

Here we derive equations for concentration-gradient-driven flow of a dilute solution in a long cylindrical pore (Fig.~\ref{Sfig:cylinder}) assuming the same governing equations used in the main text for flow through a circular aperture in an ultrathin planar membrane. We assume that the pore length $L$ is much larger than the pore radius ($L \gg a$), so that entrance effects can be ignored. The solute--wall interaction potential $U(r)$ in this case is only a function of the radial coordinate $r$. 

Since $L \gg a$, the radial solute flux will be small relative to the axial solute flux ($0 \approx j_r \ll j_z$), so from Eq.~\eqref{eq:delj-lowPe} of the main paper,
\begin{equation}
-\frac{\partial c}{\partial r} - \frac{c}{k_\mathrm{B}T}\frac{\partial U}{\partial r} = 0, 
\label{Seq:jr}
\end{equation}
which can be integrated to give
\begin{equation}
c = c_\mathrm{s}(z)e^{-U/k_\mathrm{B}T}, 
\label{Seq:ccyl}
\end{equation} 
where $c_\mathrm{s}(z)$ is the solute concentration where $U=0$. Similarly, the radial solution velocity will be  negligible compared with the axial solution velocity ($0\approx u_r \ll u_z$), so from the $r$-component of Eq.~\eqref{eq:stokes} of the main paper,
\begin{equation}
\frac{\partial p}{\partial r} + c\frac{\partial U}{\partial r} \approx 0, 
\label{Seq:stokesr}
\end{equation} 
which can be integrated to give the solution pressure as
\begin{equation}
p = \pinf + \kB T\cs(z)\left(e^{-U(r)/\kB T}-1\right),
\label{Seq:pcyl}
\end{equation} 
where $\pinf$, a constant, is the pressure where $U = 0$. Inserting the $z$-derivative of Eq.~\eqref{Seq:pcyl} into the $z$-component of Eq.~\eqref{eq:stokes} in the main paper and integrating twice using the no-slip boundary condition ($\bm{u} = 0$) at the pore surface gives the axial velocity as
\begin{equation}
u_z(r) = -\frac{\kB T}{\eta}\frac{\rmd c_\mathrm{s}}{\rmd z}\int_r^a \frac{\mathrm{d}r'}{r'}\int_0^{r'} \mathrm{d}r''r''\left(e^{-U(r'')/k_\mathrm{B}T}-1\right).  
\label{Seq:vz}
\end{equation}
The solution flux at any cross-section of the pore is
\begin{equation}
Q = \iint_S \rmd s\, \bm{u} \cdot \hat{\bm{n}}, 
\label{Seq:Q}
\end{equation}  
into which Eq.~\eqref{Seq:vz} can be inserted (with unit normal vector $\hat{\bm{n}} = \hat{\bm{z}}$) to give 
\begin{equation}
Q = -\frac{\pi\kB T}{2\eta}\frac{\rmd\cs}{\rmd z}\int_0^a \mathrm{d}rr(a^2-r^2)\left(e^{-U(r)/\kB T}-1\right). 
\label{Seq:Qcyl}
\end{equation} 
If $\Dc$ is the change in $\cs$ over the length of the pore, then using the fact that the flux $Q$ is the same for any cross-section along the pore gives
\begin{equation}
Q = -\frac{\pi\kB T}{2\eta}\frac{\Dc}{L}\int_0^a \mathrm{d}rr(a^2-r^2)\left(e^{-U(r)/\kB T}-1\right). 
\label{Seq:Qcyl2}
\end{equation} 
For the step-function interaction potential (Eq.~\eqref{eq:Ustep} of the main text, with $d = r$) and an interaction range smaller than the pore radius ($\lambda \leq a$), from Eq.~\eqref{Seq:Qcyl2} the solution flux is
\begin{equation}
\Qstep = -\frac{\pi\kB T}{8\eta}\frac{\Dc}{L}(\alpha - 1)(2a-\lambda)^2\lambda^2, \quad \lambda \leq a. 
\label{Seq:Qcylstep}
\end{equation} 
On the other hand, for an interaction range much larger than the pore radius ($\lambda \gg a$), $e^{-U(d)/\kB T}$ can be approximated as a constant, which from Eq.~\eqref{Seq:Qcyl2} gives the solution flux as 
\begin{equation}
\Qstep \simeq -\frac{\pi\kB T}{8\eta}\frac{\Dc}{L}(\alpha - 1)a^4, \quad \lambda \gg a.  
\label{Seq:Qcylstep_big}
\end{equation} 
The solute flux density may be written, using Eq.~\eqref{eq:delj-lowPe} of the main text  and Eq.~\eqref{Seq:ccyl}, as 
\begin{equation}
\bm{j} = -D e^{-U/\kB T}\frac{\rmd \cs}{\rmd z}\hat{\bm{z}},
\label{Seq:jcyl}
\end{equation}
which can be integrated over the pore cross-section using Eq.~\eqref{eq:J} of the main text to give the total solute flux as
\begin{equation}
J = -2\pi D\frac{\Dc}{L} \int_0^a \mathrm{d}rre^{-U/\kB T},  
\label{Seq:Jcyl}
\end{equation}
using the fact that the flux $J$ is conserved for any cross-section along the pore. For the step-function interaction potential (Eq.~\eqref{eq:Ustep} of the main text), the solute flux reduces to
\begin{equation}
\Jstep = 
-\pi D \frac{\Dc}{L}\left[a^2+ (\alpha - 1) \lambda(2a-\lambda)\right],\quad   \lambda \leq a
\label{Seq:Jcylstep}
\end{equation}
and
\begin{equation}
\Jstep \simeq  
-\pi D\frac{\Dc}{L} \alpha a^2 ,\quad \lambda \gg a.
\label{Seq:Jcylstep-big}
\end{equation}

\section{Finite-element numerical simulations\label{Ssec:fem}}

\begin{figure}[tb!]
	\centering
	\includegraphics{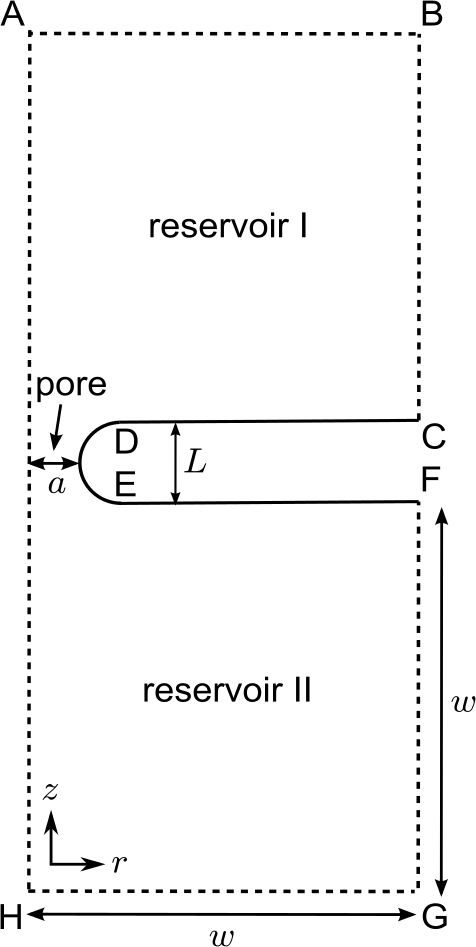}
	\caption{Schematic (not to scale) of the two-dimensional axisymmetric computational domain used in the FEM calculations. Solid lines denote solid--liquid boundaries and dashed lines liquid boundaries. The geometry has rotational symmetry about the boundary AH.
		\label{Sfig:fem_domain}
	}
\end{figure}

The continuum hydrodynamic flow equations (Eqs.~\eqref{eq:stokes}--\eqref{eq:delu} of the main paper) were solved using finite-element method (FEM) simulations with COMSOL Multiphysics version 4.3a \cite{Comsol4.3a} for a thin planar membrane containing a circular aperture of radius $a$ connecting two large cylindrical fluid reservoirs (Fig.~\ref{Sfig:fem_domain}). The equations were solved using a fully coupled solver, which is a damped version of Newton's method. The damping option used to achieve convergence was ``Automatic highly nonlinear (Newton)''. The PARDISO direct solver was used. 

The solute--membrane interaction potential was modelled using Eq.~\eqref{eq:Ufem} of the main paper. To avoid discontinuities in the potential, the membrane was given a finite thickness $L$ and the membrane surface between points D and E in Fig.~\ref{Sfig:fem_domain} was given a finite radius of curvature of $L/2$. The membrane thickness was less than 1/5 of the range of the solute--membrane potential in all cases and we verified that the measured solution and solute fluxes were not sensitive to halving the membrane thickness from the value used for the results presented. A boundary layer mesh was used at all solid--liquid interfaces, with 5 boundary layers, a thickness of the first layer of 1/2 the local domain element height and a boundary layer stretching factor (mesh element growth rate) of 1.2. The maximum element size at the boundary between points D and E in Fig.~\ref{Sfig:fem_domain} was $\min[L/5,\lambda/20]$ and that at the boundary between points C and D and between E and F was $\lambda/5$, where $\lambda$ was  the solute--membrane interaction range parameter. The maximum element size in the simulation domain was $28\sigma$, where $\sigma$ was the unit of length.  We verified that the measured solution and solute fluxes did not change significantly with a finer mesh. Table~\ref{Stab:bc} lists the boundary conditions that were used to solve the equations and Table~\ref{Stab:param} lists the parameters used in the simulations.

\begin{table}[tb!]
	\centering
	\caption{Boundary conditions used to solve the continuum equations in the FEM simulations. The vector $\bm{\hat n}$ is the surface normal.}
	\begin{tabular}{lll}
		\hline
		boundary & conditions\\
		\hline
		AH & $\bm{\hat n}\cdot \nabla c = \bm{\hat n} \cdot \bm{u} = \bm{\hat n}\cdot \nabla \bm{u} = 0$ \\
		AB & $c= \cH $, $p = \pinf $ \\
		GH & $c=\cL$, $p = \pinf$ \\
		BC and FG & $\bm{\hat n}\cdot \bm{j} = \bm{\hat n}\cdot \bm{u} = \bm{\hat n}\cdot \nabla \bm{u} = 0$ \\
		CD, DE, and EF & $\bm{\hat n}\cdot \bm{j} = \bm{u} = 0$ \\
		\hline
	\end{tabular}
	\label{Stab:bc}
\end{table}

\begin{table*}[tb!]
	\centering
	\caption{Parameters used for FEM calculations. The units of length, energy, and time are $\sigma$, $\kB T$, and $\tau = \eta\sigma^3/(\kB T)$, respectively, where $T$ is the temperature and $\eta$ is the solution shear viscosity ($\sigma$ can be considered the diameter of a fluid molecule). Where a range of values is given, the parameter was fixed at the value in parentheses unless otherwise indicated.}
	\begin{tabular}{lccl}
		\hline
		quantity & symbol & unit & value\\
		\hline
		solute diffusivity & $D$ & $\sigma^2/\tau$ & \num{1} \\
		reservoir concentration & $\cinf$ & $\sigma^{-3}$ & \num{e-3}  \\
		membrane thickness & $L$ & $\sigma$ & \num{0.2}  \\
		pore radius & $a$ & $\sigma$ & \num{1}--\num{100}  (\num{10})\\
		
		solute--membrane interaction strength & $\epsilon$ & $\kB T$ & -\num{3}--\num{3} (\num{1/10} or \num{1}) \\
		solute--membrane interaction range & $\lambda$ & $\sigma$ & \num{1}--\num{10}  (\num{1}) \\
		concentration difference & $\Dc$ & $\sigma^{-3}$ & \num{e-6}--\num{3e-4} \\
		reservoir radius/length & $w$ & $\sigma$ & \num{1000} \\
		\hline
	\end{tabular}
	\label{Stab:param}
\end{table*}   

\begin{figure}
	\centering
	\begin{subfigure}[b]{0.49\textwidth}
		\includegraphics{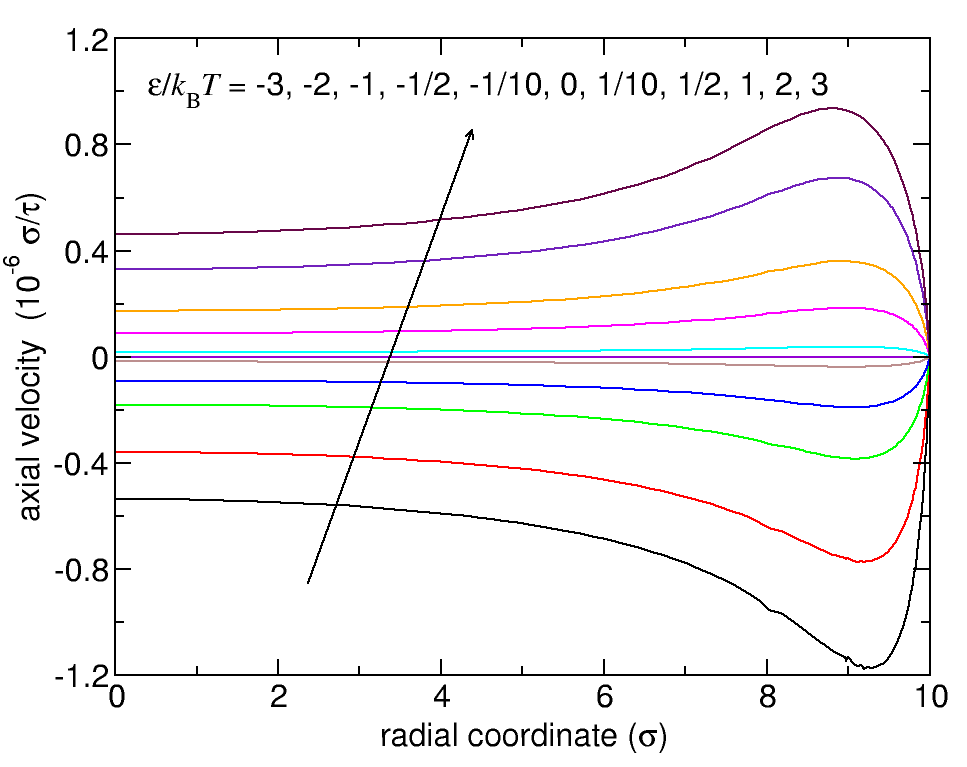}
		\caption{}
	\end{subfigure}
	\begin{subfigure}[b]{0.49\textwidth}
		\includegraphics{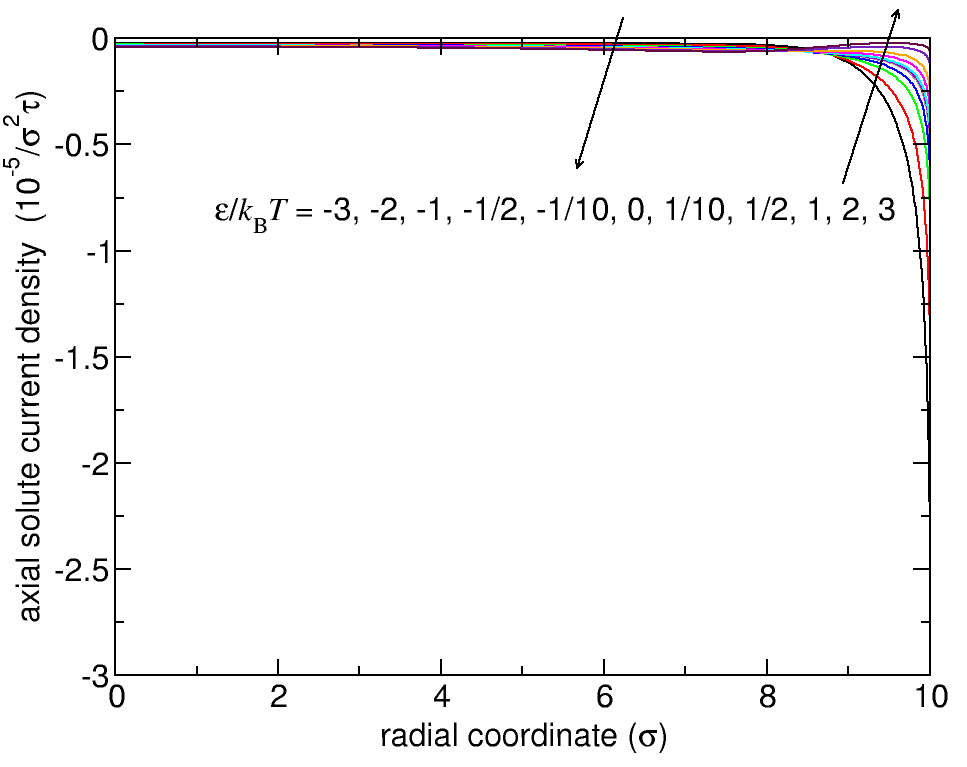}
		\caption{}
	\end{subfigure}
	\caption{(a) Axial velocity $u_z$ and (b) axial solute current density $j_z$ from FEM simulations versus the radial coordinate $r$ at the pore mouth at $z=0$ for varying solute--membrane interaction strength $\epsilon$ (for $a=10\sigma$ and $\lambda=\sigma$). The solute concentration difference $\Dc$ between the reservoirs was \num{e-5}~$\sigma^{-3}$ in all cases. (The arrows indicate the direction of increasing parameter values.)
		\label{Sfig:axial_velocity_current}
	}
\end{figure}

\begin{figure*}
	\centering
	\begin{subfigure}[b]{0.49\textwidth}
		\includegraphics[width=\textwidth]{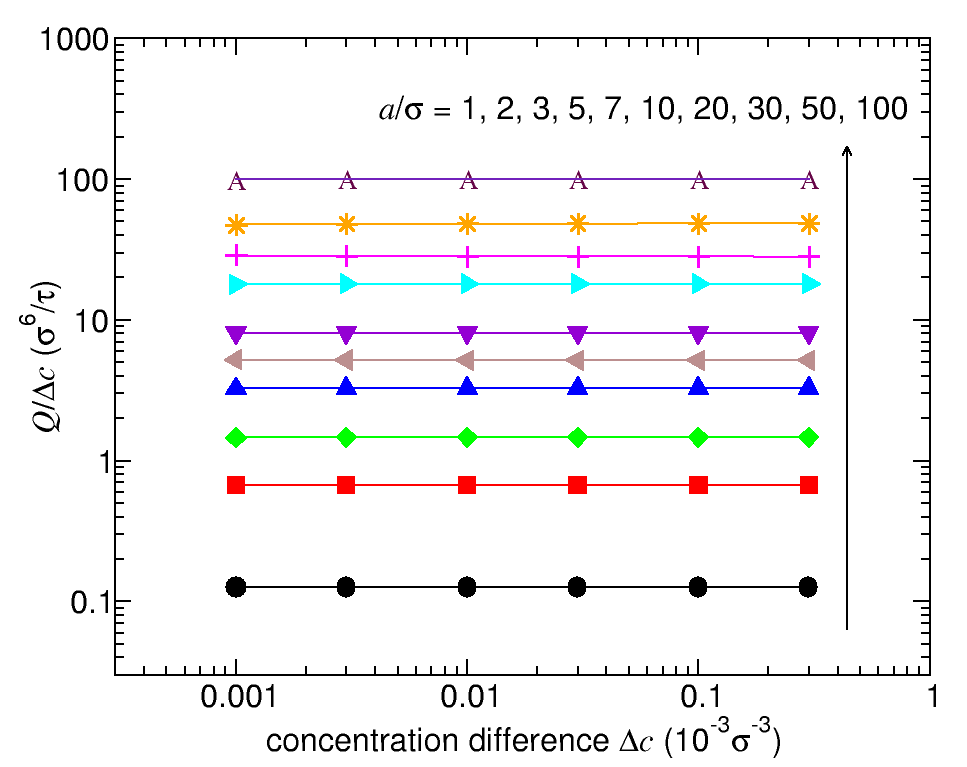}
		\caption{}
	\end{subfigure}
	\begin{subfigure}[b]{0.49\textwidth}
		\includegraphics[width=\textwidth]{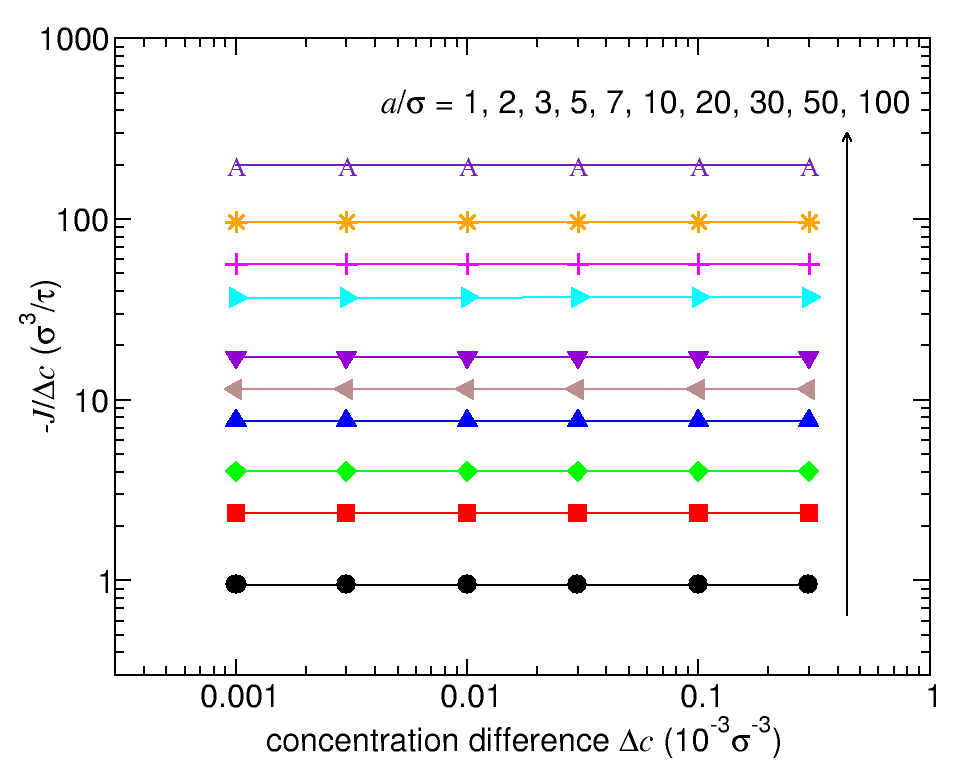}
		\caption{}
	\end{subfigure}
	\begin{subfigure}[b]{0.49\textwidth}
		\includegraphics[width=\textwidth]{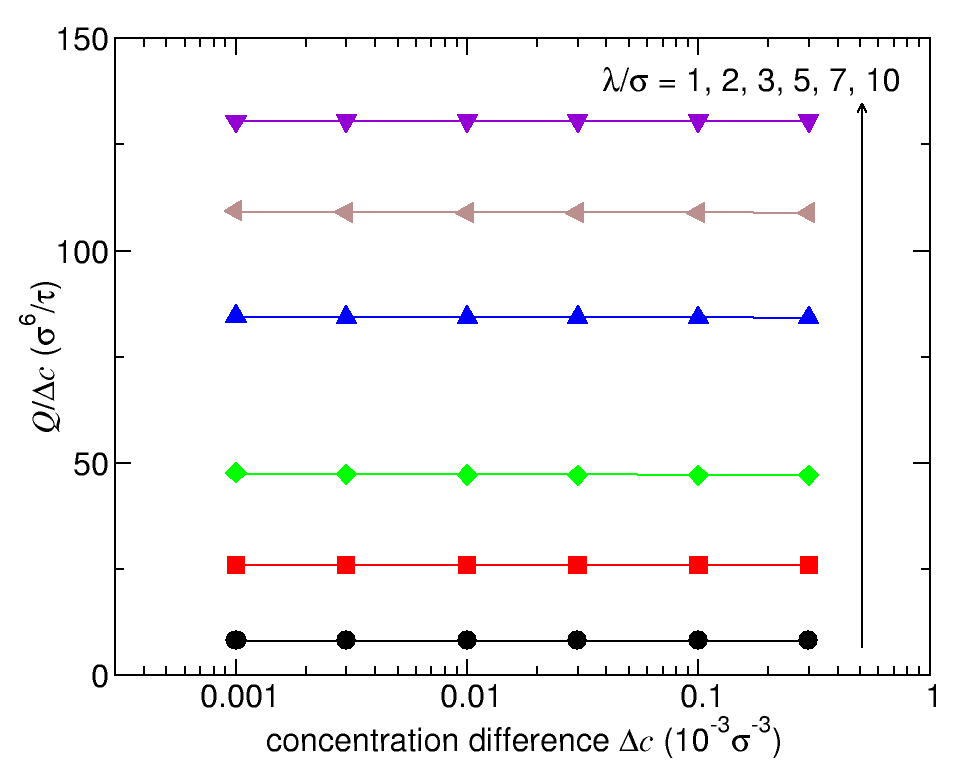}
		\caption{}
	\end{subfigure}
	\begin{subfigure}[b]{0.49\textwidth}
		\includegraphics[width=\textwidth]{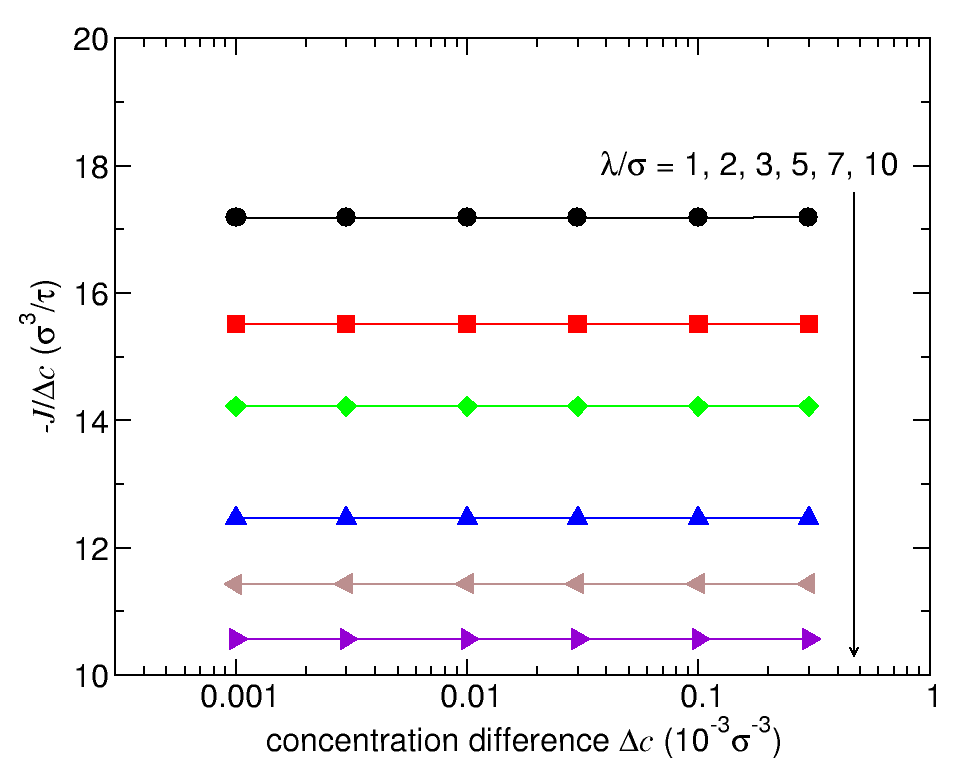}
		\caption{}
	\end{subfigure}
	\caption{
		((a), (c)) $Q/\Dc$  and ((b), (d)) $-J/\Dc$, where $Q$ is the total solution flux and $J$ is the total solute flux, versus applied concentration difference $\Dc$ from FEM simulations for varying ((a),(b)) pore radius $a$ (for $\lambda = \sigma$ and $\epsilon = \kB T$) and ((c),(d)) solute--membrane interaction range $\lambda$ (for $a = 10\sigma$ and $\epsilon = \kB T$). (The horizontal lines are linear fits to the points and the arrows indicate the direction of increasing parameter values.)
		\label{Sfig:flux_lin_resp}
	}
\end{figure*}

Several supplementary results from the FEM simulations are given in the figures below. Figure~\ref{Sfig:solute_conc} shows the simulated solute concentration $c$ and the ratio of the simulated to theoretical solute concentration, $c/c_\mathrm{theory}$, where $c_\mathrm{theory}$ is the solute concentration derived in the theory (given by Eqs.~\eqref{eq:c} and~\eqref{eq:cs} of the main paper), near the pore entrance for an intermediate solute concentration difference between the reservoirs (similiar results were obtained for the range of parameters simulated, with the maximum deviation  between $c$ and $c_\mathrm{theory}$ of around $\pm 4$\% observed for $\epsilon = -3\kB T$ and $\Dc = \num{3e-4}/\sigma^{3}$). These results indicate that the theory accurately predicts the nonequilibrium solute concentration in this system geometry, even for a solute--membrane interaction strength at the membrane surface several times larger than $\kB T$. Figure~\ref{Sfig:axial_velocity_current} gives illustrative examples of the simulated axial velocity $u_z$ and axial solute current density $j_z$ as a function of the radial coordinate $r$ at the pore mouth at $z=0$, showing in particular the effect of varying the solute--membrane interaction strength $\epsilon$.  Figure~\ref{Sfig:flux_lin_resp} shows the simulated total solution flux and total solute flux divided by the concentration difference between the reservoirs as a function of the concentration difference, demonstrating that both fluxes varied linearly with the applied concentration difference for all of simulated parameters. (Results are not shown for varying pore radius $a$ or varying solute--membrane interaction range $\lambda$ with the solute--membrane interaction strength fixed at $\epsilon = \kB T/10$, but they are qualitatively similar to those for $\epsilon = \kB T$.)  We have verified that the diffusive solute flux (i.e. not including the contribution to the flux from solute advection) in all the simulations was essentially the same as the total solute flux, demonstrating that the simulations were all at low Pecl\'et number.

\begin{figure*}
	\centering
	\begin{subfigure}[b]{0.49\textwidth}
		\includegraphics[width=\textwidth]{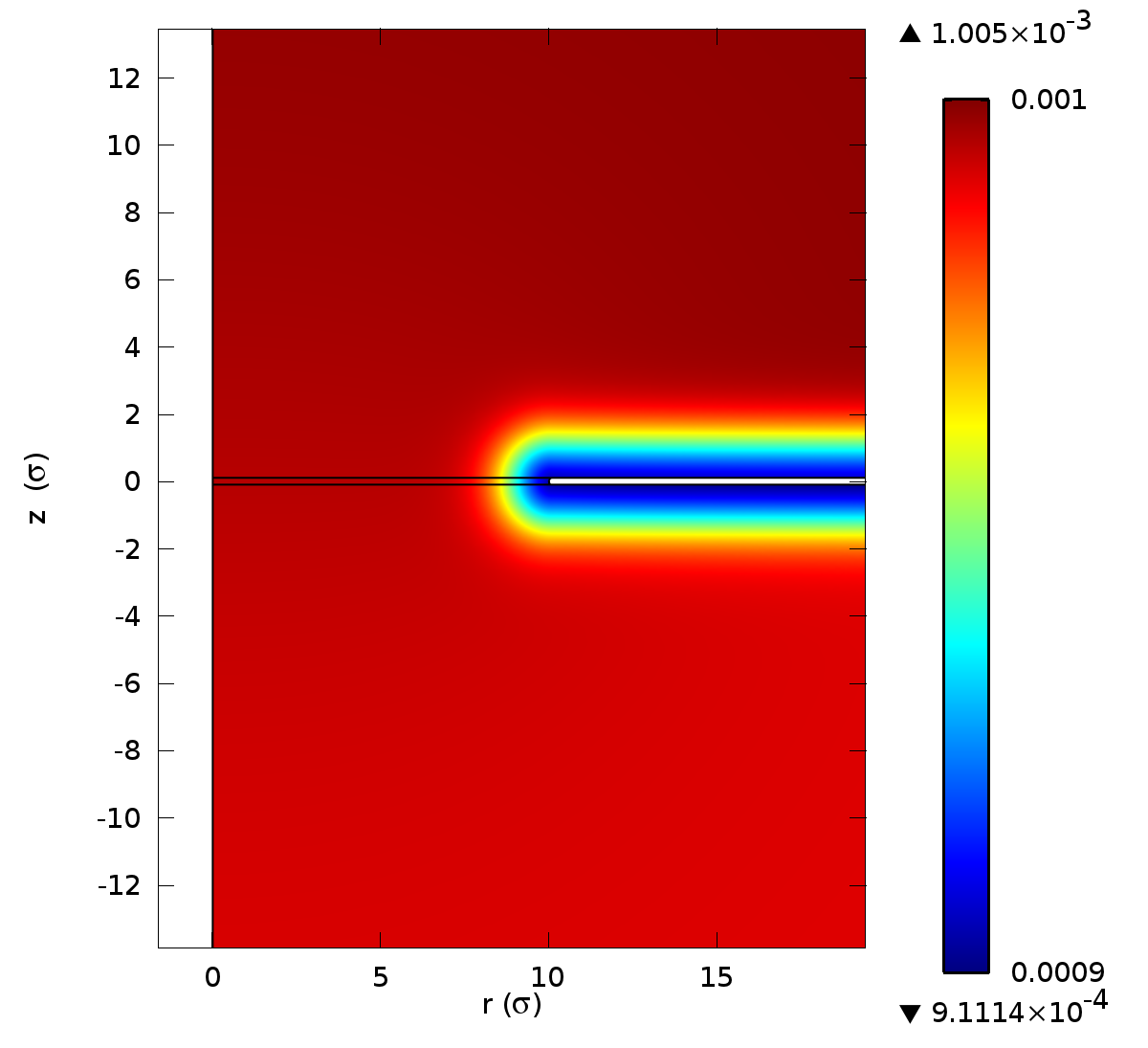}
		\caption{}
	\end{subfigure}
	\begin{subfigure}[b]{0.49\textwidth}
		\includegraphics[width=\textwidth]{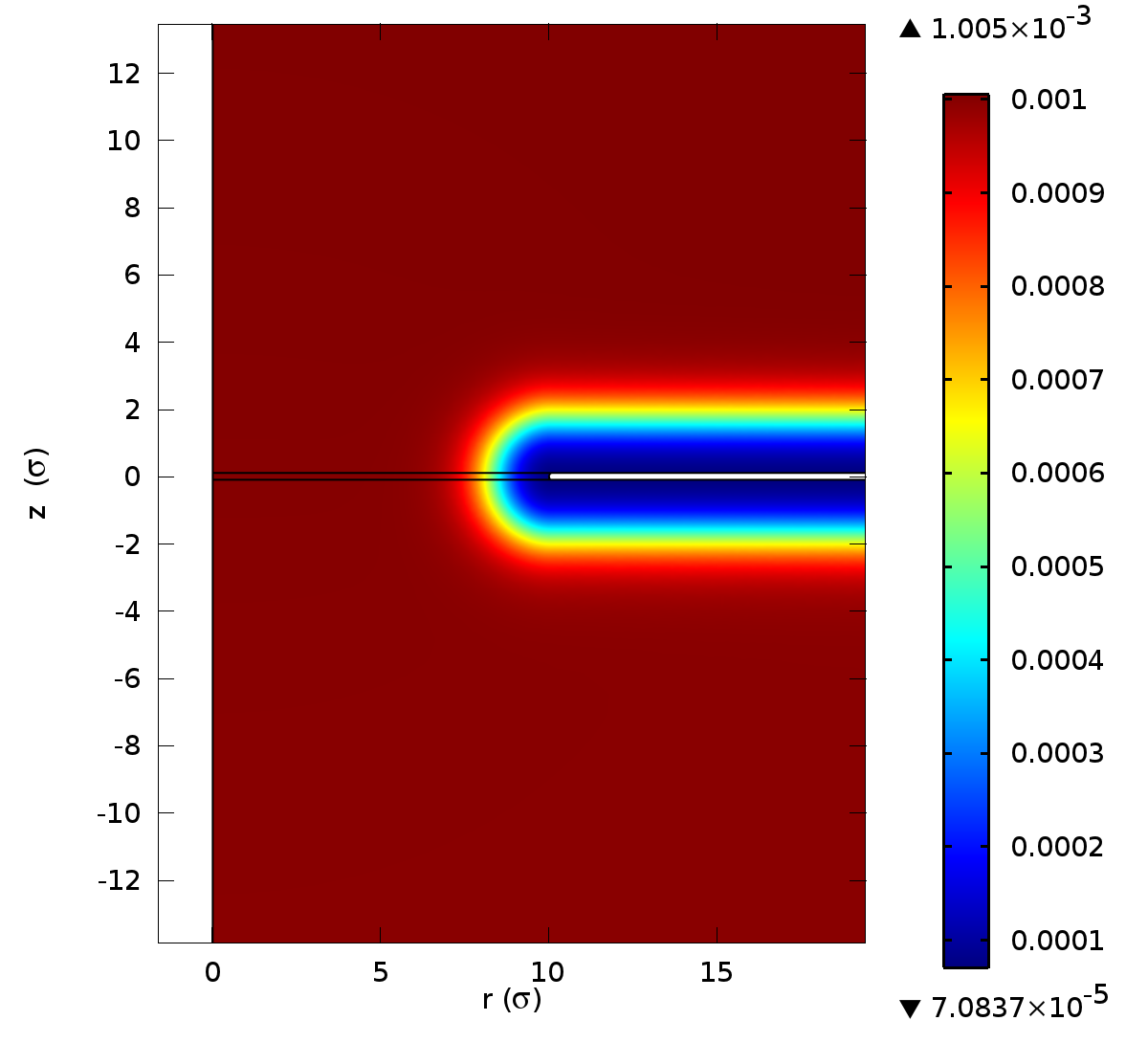}
		\caption{}
	\end{subfigure}
	\begin{subfigure}[b]{0.49\textwidth}
		\includegraphics[width=\textwidth]{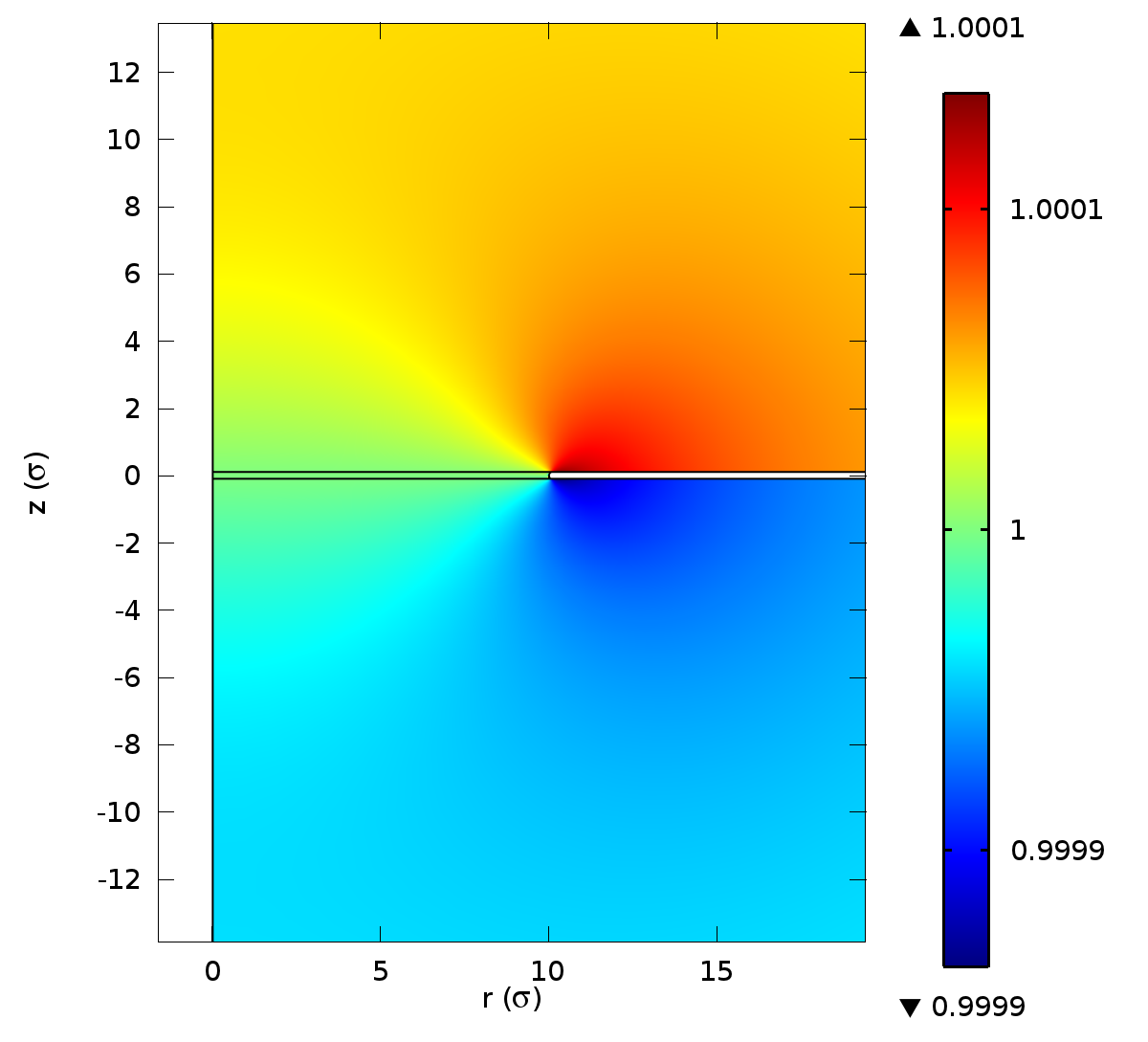}
		\caption{}
	\end{subfigure}
	\begin{subfigure}[b]{0.49\textwidth}
		\includegraphics[width=\textwidth]{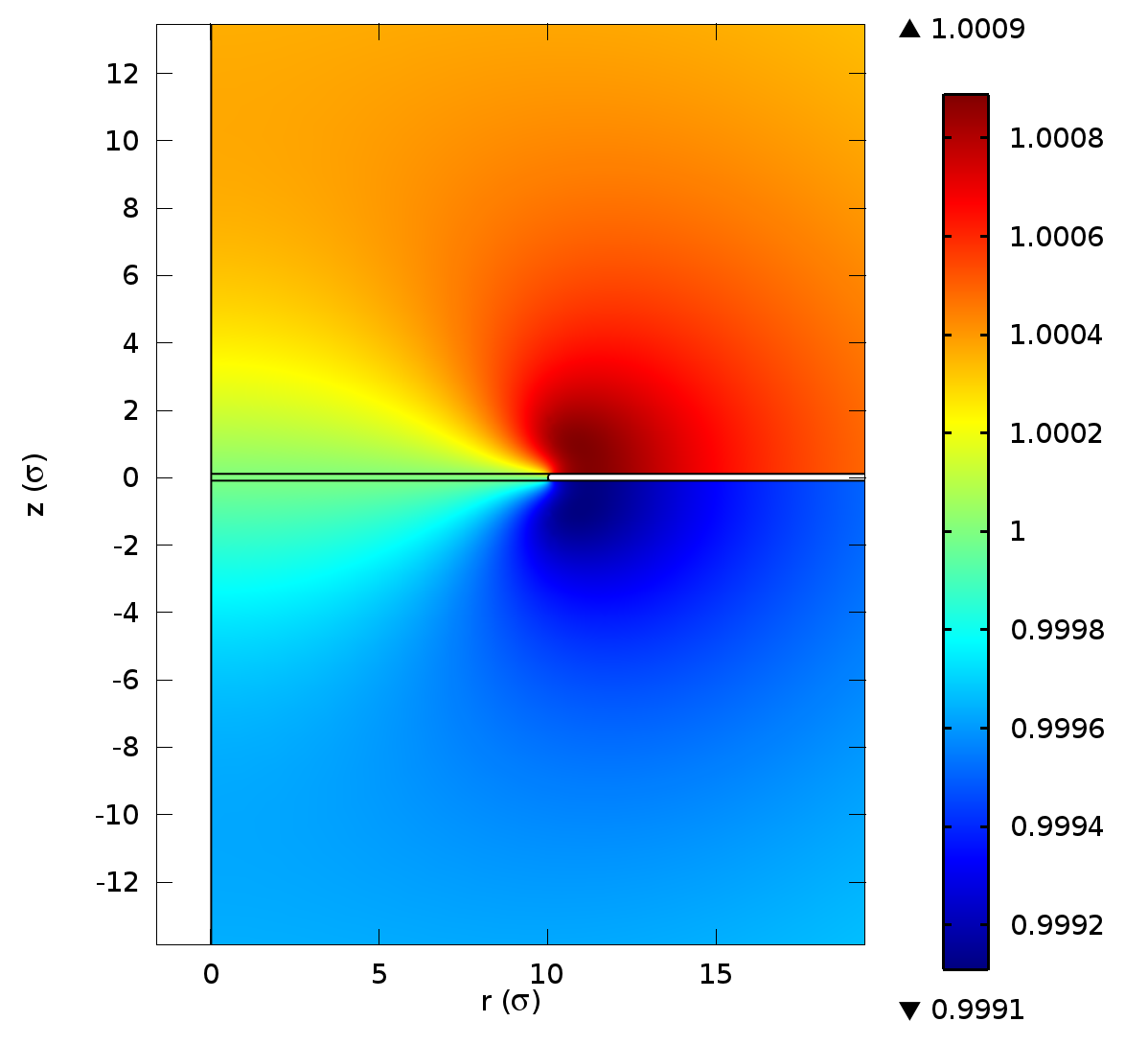}
		\caption{}
	\end{subfigure}
	
	\caption{2D plots of the ((a), (b)) solute concentration $c$ (in units of $\sigma^{-3}$) from FEM simulations  and ((c), (d)) $c/c_\mathrm{theory}$, where $c_\mathrm{theory}$ is the solute concentration derived in the theory (given by Eqs.~\eqref{eq:c} and~\eqref{eq:cs} of the main paper), as a function of the $r$ and $z$ coordinates near the pore entrance for a solute--membrane interaction strength $\epsilon$ of ((a),(c)) $\kB T/10$ and ((b),(d)) $3\kB T$ (for $a=10\sigma$ and $\lambda=\sigma$).  The solute concentration difference $\Dc$ between the reservoirs was \num{e-5}~$\sigma^{-3}$ in both cases. %
		\label{Sfig:solute_conc}
	}
\end{figure*}

% If in two-column mode, this environment will change to single-column format so that long equations can be displayed. 
% Use only when necessary.
%\begin{widetext}
%$$\mbox{put long equation here}$$
%\end{widetext}

% Figures should be put into the text as floats. 
% Use the graphics or graphicx packages (distributed with LaTeX2e).
% See the LaTeX Graphics Companion by Michel Goosens, Sebastian Rahtz, and Frank Mittelbach for examples. 
%
% Here is an example of the general form of a figure:
% Fill in the caption in the braces of the \caption{} command. 
% Put the label that you will use with \ref{} command in the braces of the \label{} command.
%
% \begin{figure}
% \includegraphics{}%
% \caption{\label{}}%
% \end{figure}

% Tables may be be put in the text as floats.
% Here is an example of the general form of a table:
% Fill in the caption in the braces of the \caption{} command. Put the label
% that you will use with \ref{} command in the braces of the \label{} command.
% Insert the column specifiers (l, r, c, d, etc.) in the empty braces of the
% \begin{tabular}{} command.
%
% \begin{table}
% \caption{\label{} }
% \begin{tabular}{}
% \end{tabular}
% \end{table}

% If you have acknowledgments, this puts in the proper section head.
%\begin{acknowledgments}
% Put your acknowledgments here.
%\end{acknowledgments}

% Create the reference section using BibTeX:
\bibliography{sup.bib}

% cross reference with manuscript document
\makeatletter\@input{ms-xr.tex}\makeatother